\documentclass[12pt, draftclsnofoot, onecolumn]{IEEEtran}
\usepackage{graphicx}
\usepackage[caption=false,textfont=sf]{subfig}
\usepackage{changes}
\usepackage{amsmath}
\usepackage{graphicx}
\usepackage{multicol}
\usepackage{amssymb}
\usepackage{subfig}
\usepackage{lineno}
\usepackage{makecell}
\usepackage{indentfirst}
\usepackage{bm}
\usepackage{color}
\usepackage{cite}
\usepackage{epstopdf}
\usepackage[noend]{algpseudocode}
\usepackage{algorithmicx,algorithm}
\usepackage{amssymb}
\usepackage{xcolor}
\usepackage{tikz}
\usepackage{mathtools}

\newcommand{\tr}{\mathrm{tr}}
\newcommand{\HH}{\mathrm{H}}	
\newcommand{\TT}{\mathrm{T}}

\newtheorem{proposition}{Proposition}

\newtheorem{lemma}{Lemma}

\ifCLASSINFOpdf
\else
\fi
\usepackage{amsmath}
\begin{document}
	\title{Perceptive Mobile Network with Distributed Target Monitoring Terminals: Leaking Communication Energy for Sensing}
	\author{Lei Xie,~\IEEEmembership{Member,~IEEE,} Peilan Wang, S.H. Song,~\IEEEmembership{Senior Member,~IEEE,}\\ and Khaled B. Letaief,~\IEEEmembership{Fellow,~IEEE}
		\thanks{L. Xie and S.H. Song are with Department of Electronic and Computer Engineering, the Hong Kong University of Science and Technology, Hong Kong. e-mail: ($\{$eelxie,eeshsong$\}$@ust.hk).} 
		\thanks{P. Wang is with the National Key Laboratory of Science and Technology on Communications, University of Electronic Science and Technology of China, Chengdu.} 
		\thanks{Khaled B. Letaief is with the Department of Electronic and Computer Engineering, the Hong Kong University of Science and Technology, Hong Kong, and also with Peng Cheng Laboratory, Shenzhen 518066, China (e-mail: eekhaled@ust.hk).}
	}
	\maketitle
	\begin{abstract}
		Integrated sensing and communication (ISAC) creates a platform to exploit the synergy between two powerful functionalities that have been developing separately. However, the interference management and resource allocation between sensing and communication have not been fully studied. In this paper, we consider the design of perceptive mobile networks (PMNs) by adding sensing capability to current cellular networks. To avoid the full-duplex operation, we propose the PMN with distributed target monitoring terminals (TMTs) where passive TMTs are deployed over wireless networks to locate the sensing target (ST). We jointly optimize the transmit and receive beamformers towards the communication user terminals (UEs) and the ST by alternating-optimization (AO) and prove its convergence. To reduce computation complexity and obtain physical insights, we further investigate the use of linear transceivers, including zero forcing and beam synthesis (B-syn).  Our analysis revealed interesting physical insights regarding interference management and resource allocation between sensing and communication: 1) instead of forming dedicated sensing signals, it is more efficient to redesign the communication signals for both communication and sensing purposes and ``leak'' communication energy for sensing;  2) the amount of energy leakage from one UE to the ST depends on their relative locations.
	\end{abstract}
	
	\IEEEpeerreviewmaketitle

	\section{Introduction}
	\IEEEPARstart{W}{ireless} communications systems have evolved for several generations and the recently commercialized 5G systems have partially met the needs for high data rate, reliable, and low latency communication services. However, with the development of innovative applications such as autonomous driving and industrial IoT \cite{cui2021integrating,10.1145/3447744,9143269}, future wireless systems are expected to provide new services, e.g., target tracking and environmental monitoring \cite{6504845,9296833}. To this end, the recently proposed integrated sensing and communication (ISAC) framework provides a promising platform to integrate sensing capability in communication systems \cite{liu2021integrated}. The use of millimeter-wave (mmWave) in 5G and beyond further facilitates the integration between radar and communication systems with possible hardware and software sharing between two functionalities. In this case, there have been some interesting developments in three different areas, i.e., dual-functional radar-communication (DFRC) \cite{8288677,8386661,8999605,9124713}, sensing-aided communication \cite{5776640}, and communication-aided sensing \cite{1211013,4268440,7814210,7347464,7485066,9093221,9205659}.
	
	The most fundamental issue for integrating sensing and communication lies in the interference management and resource allocation between the two sub-systems. The interference exists and can be handled in both the device and network levels. In the device level, full-duplex operation is a critical challenge because BSs need to serve as the transmitter and receiver simultaneously \cite{9540344}. To this end, existing works considered self-interference cancellation (SIC) \cite{bharadia2013full,sabharwal2014band,korpi2013advanced}, where analog filters based on the recursive least squares (RLS) or least mean squares (LMS) estimation are proposed.

	In the system level, there exists the interference between communication and sensing signals. To this end, some research efforts have been made for the transmit beamformer and waveform design in ISAC systems. \cite{8288677} optimized the transmit beamformers for the dual DFRC systems to achieve a target radar beam pattern while satisfying a given communication performance requirement. \cite{8386661} investigated the transmit waveform design by minimizing the multi-user interference in communication while allowing a tolerable mismatch between the designed and the desired radar beam patterns. The hybrid beamforming technique was considered in \cite{8999605} to save the energy consumption. To improve the parameter estimation performance, \cite{liu2021cramerrao} proposed to design the ISAC system by minimizing the Cram\'{e}r-Rao bound (CRB) for sensing. In \cite{8827589,9349171}, the perceptive mobile network (PMN) was studied to enable different types of sensing capabilities for wireless communications. However, the joint transmitter and receiver design for PMN systems is not available in the literature. More importantly, the interplay between sensing and communication, especially in terms of interference management and resource allocation, has not been well understood.

	In this paper, we will try to tackle the interference in the  system levels  by taking advantage of the network structure of mobile networks. Specifically, we first propose the PMN with distributed target monitoring terminals (PMN-TMT), where TMTs are deployed to add sensing capability to mobile networks. In the proposed PMN-TMT, the base stations (BSs) will not only serve communication but also work as the radar transmitters. In particular, BSs will transmit/receive communication signals and also send sounding signals to the sensing target (ST). However, to avoid the full-duplex operation and reduce interference between sensing and communication, the radar signal estimation task will be taken over by the TMTs, which are passive sensing terminals deployed as IoT devices for sensing and monitoring tasks \cite{9143269,7845396}. They can be deployed on the BSs, but normally will be spread around the BSs to provide additional angles for sensing and monitoring purposes. TMTs are connected to the BSs through high capacity links and form the sensing network.

	With the proposed PMN-TMT, we will then  consider the joint design of transmit and receive beamformers. Due to the different nature of sensing and communication systems, the PMN-TMT has some unique characteristics: 1) Sensing signal, if not handled properly, will cause interference to the communication receivers; 2) Communication signals can be utilized for sensing purpose, given they are known by the BSs; and 3) Communication signals will be reflected by the environment, creating the clutter (interference) for sensing, which may lead to false alarm/estimation. As a result, how to manage the interference and allocate resources between sensing and communication are two of the most important questions to be addressed.  In this paper, we will jointly design the transmitter and receiver to optimize the sensing and communication performance. This problem is first solved by an alternating optimization (AO) framework, whose convergence proof is also given.  However, the complexity of the AO-based design is high.  To reduce the complexity and obtain physical insights regarding the interference management and resource allocation between the two sub-systems, we further derive linear transceiver structures and compare their performance with that of the AO-based solution.     
	
	The contributions of this paper can be summarized as follows:
	\begin{enumerate}
		\item We propose a novel ISAC framework, i.e., PMN-TMT, where passive TMTs are deployed over traditional mobile networks to locate the ST. The distributed multi-antenna TMTs save the need for full-duplex operation, bring macro-diversity for target detection, and provide spatial freedom for interference management and environment sensing.
		
		\item We jointly design the transmitter and receiver by maximizing the weighted average of sensing and communication performance. To address the non-convex optimization problem, we transfer the fractional programming to a parametric square-root subtractive-form problem by exploiting the quadratic transform technique \cite{8314727}. Then, we propose an AO-based framework to iteratively optimize the transmit and receive beamformers  and  prove the convergence of the proposed algorithm.
		
		\item  We derive linear transceiver structures, including the zero-forcing (ZF) and beam synthesis (B-syn) transmitter, and the minimum variance distortionless response (MVDR) receiver. These linear transceivers not only reduce the computation complexity but also provide interesting physical insights: (1) ``Leaking'' energy from communication signals to the ST is more efficient than forming a dedicated sensing signal; and (2) the amount of energy leaked from one UE to the ST depends on their channel correlation, which is determined by their locations. 
	\end{enumerate}
	
	The remainder of this paper is organized as follows. Section II introduces the system model of the proposed PMN-TMT. Section III formulates the problem and provides the AO-based joint transceiver design algorithm, together with its convergence proof. The sub-optimal linear transceiver structures are derived in Section IV, where several interesting physical insights regarding the interference management and resource allocation between sensing and communication are also revealed. Section V provides  simulation results to illustrate the performance of the proposed methods  and  Section VI concludes the paper.
	
	\section{System Model}
	
	\subsection{Perceptive Mobile Network with Distributed Target Monitoring Terminals}
\begin{figure}[!t]
		\centering
		\includegraphics[width=3.2in]{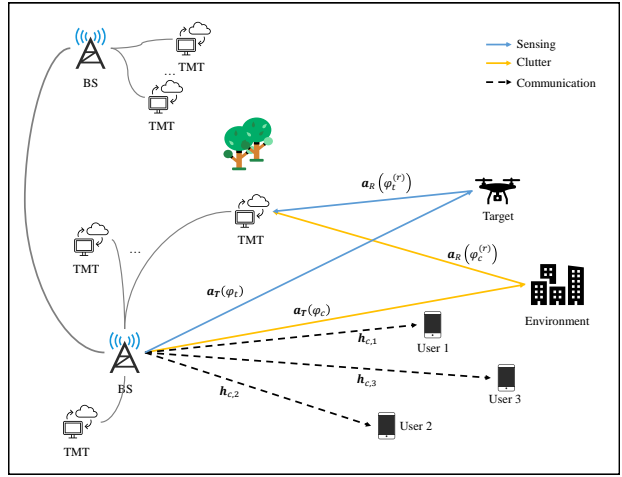}
		\caption{Illustration of PMN-TMT.}
		\label{sys1}
	\end{figure}

	PMN represents a promising framework to integrate radar sensing in wireless communications networks. However, due to the different nature of communication and sensing functionalities, a fully integrated system with dual-functionality will face many challenging issues such as the full-duplex operation. In this paper, to release such demanding requirements, we propose the PMN-TMT, as illustrated in Fig. \ref{sys1}. The PMN-TMT can be implemented by adding another layer of passive TMTs over the traditional cellular networks. In particular, TMTs are passive nodes with only perception functionalities, including radar, vision, and other sensing capabilities \cite{9143269,7845396}. They are distributed in a target area and connected with the base stations (BSs) through low latency links.   
	
	In the proposed PMN-TMT, the communication between the BSs and the UEs is achieved in the same way as traditional cellular networks. To perform radar sensing, the sounding signal is generated by the BSs during the downlink communication period. 
	To avoid transmitting and receiving at the same time (full-duplex), the target sensing/detection is performed by the TMTs. The distributed TMTs not only reduce the implementation difficulty but also provide multiple angles to monitor the environment. Such design also facilitates the integration of other IoT applications in the PMN-TMT. 
	
	In this paper, we consider a simplified network with one base station (BS) and one TMT equipped with $N_{t}$ and $N_{r}$ antennas\footnote{Collaborative sensing by several TMTs will provide better sensing performance, which is left for the future work.}, respectively. The multiple antennas at the TMTs provide the spatial freedom that for environment sensing and interference rejection. This network serves $K$ single-antenna UEs and detects a ST simultaneously. Denote $\mathbf{h}_{c,k}\in \mathbb{C}^{N_{r} \times 1}$ and $\mathbf{h}_{R}\in \mathbb{C}^{N_{r} \times 1}$ as the channels from the BS to the $k$th UE and the ST, respectively. To serve both communication and sensing, the BS transmits $N_s$ data streams by $N_{t}$ transmit antennas. The transmit signal  $\mathbf{s}\in\mathbb{C}^{N_s \times 1}$ consists of $K$ data streams for the UEs and $N_s-K$ \footnote{Here we assume that the number of UEs $K$ is no more than $N_s$.} data streams for sensing. Without loss of generality, the transmitted symbol vector is given by
	\begin{equation}
		\begin{split}
			\mathbf{s}=\left[\begin{matrix}
				\mathbf{s}_c\\
				\mathbf{s}_R\\
			\end{matrix}\right] \in \mathbb{C}^{N_s\times 1},
		\end{split}
	\end{equation}
	which is assumed to be Gaussian distributed with zero means and covariance matrix $\mathbf{I}$. Each entry of $\mathbf{s}$ corresponds to a single-carrier waveform, where $\mathbf{s}_c$ and $\mathbf{s}_R$ denote the symbols for communication and sensing, respectively. Specially, $s_k$, i.e., the $k$th entry of $\mathbf{s}_c$, denotes the symbol transmitted to the $k$-th UE.

	In this paper, we consider the widely used Saleh-Valenzuela (SV) model to characterize the sparse nature of mmWave channels \cite{7445130,6834753,9234098}.
	Suppose uniform linear arrays (ULAs) are employed at the BS. The BS-UE channel $\mathbf{h}_{c,k}$ can then be given by \cite{9234098}
	\begin{equation}
		\begin{split}
			\mathbf{h}_{c,k}=\sqrt{\frac{N_{t}}{N_{p}}} \sum_{i=1}^{N_{p}} \beta_{k,i}^{(t)} \mathbf{a}_{T} (\phi_{k,i}^{(t)}), 
		\end{split}
		\label{channelmodel}
	\end{equation}
	where $N_{p}$ denotes the number of paths between the BS and the UE. $\mathbf{a}_{T}(\phi)$  represents  the steering vector of the BS with $||\mathbf{a}_{T}(\phi)||^2=1$ and $\phi_{k,i}^{(t)}$ is the angle-of-departure (AOD) from the BS to the $k$th UE in the $i$th path.  $\beta_{k,i}^{(t)}$  denotes the path gain in the $i$th path of the corresponding channel.  The BS-ST and ST-TMT channels can be modeled similarly. 
	
	\subsection{Communication Signal Model}
	The signal received at the $k$th UE is expressed as \cite{7397861}
	\begin{equation}
		\begin{split}
			y_{c,k}&=\mathbf{h}_{c,k}^\HH \mathbf{F}  \mathbf{s}+n_c,\\
		\end{split}
	\end{equation}
	where  
	\begin{equation}
		\begin{split}
			\mathbf{F}=\left[ \underbrace{\mathbf{f}_{c,1}, \cdots,\mathbf{f}_{c,K}}_{\mathbf{F}_c:\;\text{Communications}},\underbrace{\mathbf{f}_{R,1},\cdots,\mathbf{f}_{R,N_s-K}}_{\mathbf{F}_R:\;\text{Sensing}} \right] \in \mathbb{C}^{N_t\times N_{s}}
		\end{split}
	\end{equation}
	denotes the precoder matrix and $n_c$ represents the additive white Gaussian noise (AWGN) with zero mean and variance $\sigma_c^2$. 
	In general, it  may not be optimal to use only $K$ data streams for $K$ multiple-antenna communication users. More streams will provide additional freedom in achieving a good tradeoff between communication and sensing. In this paper, we consider single-antenna UEs, and thus we assume that $K$ streams of communication signals are transmitted.
	Note that both communication and sensing signals may impose interference to the UEs. Thus, the received signal-to-interference-plus-noise-ratio (SINR) at the $k$th UE is expressed as
	\begin{equation}
		\begin{split}
			\gamma_k \left(\mathbf{F} \right)=\dfrac{\left\vert \mathbf{h}_{c,k}^\HH \mathbf{f}_{c,k}  \right\vert^2}{
				\underbrace{	\sum_{i\neq k}\left|\mathbf{h}_{c,k}^\HH\mathbf{f}_{c,i} \right|^2}_{\text{Multi-user Interference}}
				+\underbrace{	\left\Vert\mathbf{h}_{c,k}^\HH\mathbf{F}_R \right\Vert^2}_{\text{Sensing}}
				+\sigma_c^2} . 
		\end{split}
		\label{gammak}
	\end{equation} 
	
	\subsection{Radar Signal Model}
	The sensing signal is regarded as interference by communication systems, but communication signals can be used for sensing because the transmitted communication waveform is known by the TMT. Assuming that the transmit waveform is narrow-band and the propagation path is a  line of sight (LoS) path, the base-band signal at a point-like target can be given by  \cite{9124713}
	\begin{equation}
		\begin{split}
			y_t=\mathbf{a}_{T}^\HH(\phi_t)\mathbf{F} \mathbf{s},
		\end{split}
	\end{equation} 
	where $\phi_t$ denotes the AOD from the BS to the ST. Note that the back-scattered echos from the environment will impose interference to the TMT.
	
	Target detection is a binary hypothesis testing problem, where hypotheses $\mathcal{H}_0$ and $\mathcal{H}_1$ correspond to the absence and presence of the ST, respectively. Besides the echo back-scattered from the ST, the TMT will also receive the echoes from the environment, which is known as the clutter \cite{6081358,303737,9052470}. The echo signal received by the TMT in the range-Doppler cell under test (CUT) \cite{6081358,303737,9052470} is expressed as 
	\begin{equation} \label{assuptionmodel}
		\left\{
		\begin{array}{ll}
			\mathcal{H}_0: &\mathbf{y}_{R,0}=\sum\limits_{l=1}^{L}\epsilon_{c,l} \mathbf{A}_{c,l} \mathbf{F} \mathbf{s}+\mathbf{n}_{R},\\
			\mathcal{H}_1: &\mathbf{y}_{R}=\epsilon_s\mathbf{A}_{R}  \mathbf{F} \mathbf{s} + \sum\limits_{l=1}^{L}\epsilon_{c,l} \mathbf{A}_{c,l} \mathbf{F} \mathbf{s}+\mathbf{n}_{R},\\
		\end{array}
		\right.
	\end{equation} 
	where $L$ denotes the number of clutter patches. The response matrices of the clutter patches and the target \cite{liu2021cramerrao} are respectively given as
	\[\mathbf{A}_{c,l}=\mathbf{a}_R(\phi_{c,l}^{(r)}) \mathbf{a}_{T}^\HH(\phi_{c,l}),\mathbf{A}_{R}=\mathbf{a}_{R}(\phi_t^{(r)}) \mathbf{a}_{T}^\HH(\phi_t),\]
	where $\mathbf{a}_{R}(\phi)$ denotes the steering vector of the TMT with $||\mathbf{a}_{R}(\phi)||^2=1$. $\phi_t^{(r)}$ and $\phi_{c,l}^{(r)}$ represent the angle-of-arrival (AOA) from the ST and $l$th clutter patch to the TMT, respectively. $\phi_{c,l}$ denotes the AOD of the $l$th clutter patch with respect to (w.r.t.) the BS. $\epsilon_s$ and $\epsilon_{c,l}$ represent the complex gains of the ST-TMT channel and the channel between the $l$th clutter patch and the TMT, respectively,  which depend on the gain of the matched filtering, the gain of emission patterns, the propagation loss, and the target radar cross section (RCS) \cite{303737,9052470}. They are assumed to be zero mean Gaussian random variables with variance $\sigma_t^2$ and $\sigma_{c,l}^2$, respectively \cite{303737,9052470}. $\mathbf{n}_{R} $ is modeled as AWGN with zero means and covariance matrix $\sigma_n^2 \mathbf{I}$.
	
	In practical applications, the TMT is deployed on a smart manufacturing or industrial IOT device whose power is limited \cite{9143269,7845396}. To reduce hardware cost and improve the overall energy efficiency (EE) of the TMT, we consider the hybrid beamforming structure. The baseband signal is processed by an analog baseband combiner $\mathbf{W}_{RF}\in \mathbb{C}^{N_r \times N_{RF}^r}$, a digital baseband combiner $\mathbf{W}_{BB}\in \mathbb{C}^{N_{RF}^r \times N_s}$, and a detection filter $\mathbf{w}_d\in \mathbb{C}^{N_{s} \times 1}$ with the  output
	\begin{equation}\label{xo}
		\begin{split}
			&x_{o}=\mathbf{w}_d^\HH \mathbf{W}_{BB}^\HH \mathbf{W}_{RF}^\HH \mathbf{y}_R.
		\end{split}
	\end{equation}
	Denote $\mathbf{w}=\mathbf{W}_{BB} \mathbf{w}_d \in \mathbb{C}^{N_{RF}^r \times 1}$ as the effective digital baseband processor.

	Different performance metrics can be selected for different radar applications, e.g., the Cram\'{e}r-Rao bound for angle estimation \cite{9652071}. In this paper, we focus on target detection in the clutter-presence scenario, where the signal-dependent clutter is generally  much stronger than the signal component. In this case, the detection probability is directly related to the signal-to-clutter-and-noise-ratio (SCNR) \cite{9052470}, which is widely used to measure the ability of a radar system in rejecting the clutter \cite{303737,9052470}. Therefore, we choose SCNR as the performance metric \footnote{The objective of this paper is to probe whether a target exists at a given direction. In order to scan a large area, one option is to design the transceiver for monitoring over a continuous range of directions.}, i.e.,
		\begin{equation}\label{SCNR}\nonumber
			\begin{split}
				&\text{SCNR}(\mathbf{w},\mathbf{W}_{RF},\mathbf{F})=\frac{\mathcal{P}_S(\mathbf{w},\mathbf{W}_{RF},\mathbf{F})}{\mathcal{P}_{Q}(\mathbf{w},\mathbf{W}_{RF},\mathbf{F})}=\frac{\mathbb{E}\left( \left\Vert 
					\epsilon_s \mathbf{w}^\HH \mathbf{W}_{RF}^\HH\mathbf{A}_{R} \mathbf{F}\mathbf{s}
					\right\Vert^2  \right)
				}{
					\mathbb{E}\left( \left\Vert 
					\sum\limits_{l=1}^{L}\epsilon_{c,l}\mathbf{w}^\HH \mathbf{W}_{RF}^\HH\mathbf{A}_{c,l} \mathbf{F}\mathbf{s}+ \mathbf{w}^\HH \mathbf{W}_{RF}^\HH\mathbf{n}_{R}
					\right\Vert^2  \right)},\\
			\end{split}
		\end{equation}
	where	\begin{equation}
		\begin{split}
			\mathcal{P}_S(\mathbf{w},\mathbf{W}_{RF},\mathbf{F})
			&\triangleq \sigma_t^2 \mathbf{w}^\HH\mathbf{W}_{RF}^\HH\mathbf{A}_{R}\mathbf{F}\mathbf{F}^\HH \mathbf{A}_{R}^\HH \mathbf{W}_{RF}\mathbf{w},\\
			\mathcal{P}_{Q}(\mathbf{w},\mathbf{W}_{RF},\mathbf{F})
			&\triangleq	\underbrace{ \sum\limits_{l=1}^{L}\sigma_{c,l}^2 \mathbf{w}^\HH \mathbf{W}_{RF}^\HH\mathbf{A}_{c,l}\mathbf{F}\mathbf{F}^\HH\mathbf{A}_{c,l}^\HH \mathbf{W}_{RF}\mathbf{w}}_{\text{Clutter}}+\underbrace{\sigma_n^2 \mathbf{w}^\HH\mathbf{W}_{RF}^\HH\mathbf{W}_{RF}\mathbf{w} }_{\text{Noise}}.
		\end{split}
	\end{equation}
denote the power of received signal and that of the clutter plus noise, respectively.

	\section{Joint Communication and Sensing Design}
	The communication performance can be measured by the SINR at the UEs, whereas the sensing performance depends on SCNR. In this paper, we attempt to jointly optimize the receive filter $\mathbf{w}$, the analog combiner $\mathbf{W}_{RF}$, and the precoder $\mathbf{F}$ to maximize the weighted sum of the SCNR and the worst case SINR simultaneously. Concretely, the optimization problem is formulated as
	\begin{subequations}
		\begin{align}
			\max_{\mathbf{w},\mathbf{W}_{RF},\mathbf{F}}\quad &\mathcal{L}(\mathbf{w},\mathbf{W}_{RF},\mathbf{F}) \notag\\
			s.t. \quad \quad 	&\left\Vert \mathbf{w} \right\Vert^2 \leq 1  \label{cont04}\\
			&\left\Vert \mathbf{F} \right\Vert_F^2 \leq P  \label{cont05}\\
			&\mathbf{W}_{RF}\in \mathcal{M}^{N_{r}\times N_{RF}^r}, \label{cont06}
		\end{align}	
		\label{Problem0}	
	\end{subequations}
	where
	\begin{equation}
		\begin{split}\nonumber
			&\mathcal{L}(\mathbf{w},\mathbf{W}_{RF},\mathbf{F})=\kappa_r\text{SCNR}(\mathbf{w},\mathbf{W}_{RF},\mathbf{F})+\kappa_c \min_{k\in [1,K]}
			\gamma_k(\mathbf{F}),
		\end{split}		
	\end{equation}
	with $\kappa_r \in [0,1]$ and $\kappa_c = 1-\kappa_r$ denoting the weighting coefficients for the sensing and communication, respectively. The feasible set of the analog combiner is given by $\mathcal{M}^{M\times N}=\{\mathbf{X}\in \mathcal{C}^{M\times N} \big| \ |\mathbf{X}(i,j)|=1, i=1,\cdots,M, j=1,\cdots,N \}.$
	Note that (\ref{cont04}) forces the receiver to have a unit norm and (\ref{cont05}) limits the power of the transmitter. (\ref{cont06}) represents the unit modulus constraint as the analog precoders are implemented by phase shifters. The knowledge of $\sigma_t^2$, $\sigma_n^2$ and $\sigma_{c,l}^2, l=1,\cdots,L$ can be obtained by a cognitive paradigm \cite{https://doi.org/10.1049/iet-rsn.2014.0527,melvin2006knowledge,guerci2010cognitive} and is assumed to be known. Moreover, we assume that the channel state information (CSI), i.e., $\mathbf{h}_{c,k}$ and $\mathbf{h}_{c,k}^{(r)}$, are known. It can be observed that both the fractional objective function and constraints are non-convex, which causes the optimization problem hard to solve.
	
	Note that the problem in (\ref{Problem0}) is a multi-ratio fractional programming (FP) problem which is NP-hard. Fortunately, the objective function is continuous and has positive denominator. Thus, the FP problem can be transformed into a parametric subtractive-form problem by exploiting the Dinkelbach method \cite{barros1996new}. However, although the Dinkelbach method can be an efficient solution to those single-ratio problems with a concave numerator and convex denominator, it cannot be easily generalized to the multi-ratio problem, like (\ref{Problem0}). Also, the numerator of the objective function in (\ref{Problem0}) has a quadratic form w.r.t.  $\mathbf{w}$, $\mathbf{W}_{RF}$, and $\mathbf{F}$, and thus is non-concave. As a result, extra relaxations are needed to further relax the resulting subtractive-form objective function which may degrade the convergence and optimization performance.
	To address this issue, we reformulate (\ref{Problem0}) as a parametric subtractive-form problem by exploiting the quadratic transform technique proposed in \cite{8314727}, i.e.,
	\begin{equation}
		\begin{split}
			\max_{\substack{\mathbf{w},\mathbf{W}_{RF},\mathbf{F}, \mathbf{u}_r,u_{k}} } &	\mathcal{F}\left(\mathbf{w}^{(t)},\mathbf{W}_{RF}^{(t)},\mathbf{F}^{(t)},\mathbf{u}_r^{(t)},u_{k}^{(t)} \right)\\
			s.t. \quad\quad & (\ref{cont04}) \text{-} (\ref{cont06}),\\
		\end{split}
		\label{probfracF}
	\end{equation}
	where
	\begin{equation}
		\begin{split}
			&\mathcal{F}\left(\mathbf{w}^{(t)},\mathbf{W}_{RF}^{(t)},\mathbf{F}^{(t)},\mathbf{u}_r^{(t)},u_{k}^{(t)} \right)=\mathcal{F}_R\left(\mathbf{w}^{(t)},\mathbf{W}_{RF}^{(t)},\mathbf{F}^{(t)},\mathbf{u}_r^{(t)} \right)+\min_{k\in[1,K]}\mathcal{F}_k\left(\mathbf{F}^{(t)},u_{k}^{(t)} \right),
		\end{split}
	\end{equation}
	with
	\begin{equation}
		\begin{split}
			&\mathcal{F}_R \left(\mathbf{w},\mathbf{W}_{RF},\mathbf{F},\mathbf{u}_r  \right)  = 2 \kappa_r \Re\left( \sigma_t \mathbf{w}^\HH\mathbf{W}_{RF}^\HH\mathbf{A}_{R}\mathbf{F}\mathbf{u}_r \right) -\kappa_r \Vert \mathbf{u}_r \Vert^2  \mathcal{P}_{Q}\left(\mathbf{w},\mathbf{W}_{RF},\mathbf{F} \right),
		\end{split}
		\label{lossfunc}
	\end{equation}
	\begin{equation}
		\begin{split}
			&\mathcal{F}_k \left(\mathbf{F} ,u_{k} \right)=2\kappa_c\Re\left(u_{k}  \mathbf{h}_{c,k}^\HH \mathbf{f}_{c,k} \right)-\kappa_c|u_{k} |^2  \left(\left\Vert \mathbf{h}_{c,k}^\HH \mathbf{F}_R  \right\Vert^2+\sum_{i\neq k}\left|\mathbf{h}_{c,k}^\HH\mathbf{f}_{c,i} \right|^2
			+\sigma_c^2\right).
		\end{split}
		\label{lossfuncCk}
	\end{equation}
	Here, $\mathbf{u}_r$ and $u_{k}$ are two auxiliary complex variables. 
	To solve this problem, an iteration process based on AO is given as follows 
		\begin{subequations}\label{IterationCore}
			\begin{align}
				&\mathbf{w}^{(t+1)}= \mathop{\arg \max}_{\mathbf{w}}  \mathcal{F}_R\left(\mathbf{w}|\mathbf{W}_{RF}^{(t)},\mathbf{F}^{(t)},\mathbf{u}_r^{(t)} \right), \quad s.t. \quad (\ref{cont04}), \label{probwt1}\\
				&\mathbf{W}_{RF}^{(t+1)}= \mathop{\arg \max}_{\mathbf{W}_{RF}}  \mathcal{F}_R\left(\mathbf{W}_{RF}|\mathbf{w}^{(t+1)},\mathbf{F}^{(t)},\mathbf{u}_r^{(t)} \right), \quad s.t. \quad  (\ref{cont06}),\label{probWRFt1}\\
				&\mathbf{F}^{(t+1)} = \mathop{\arg \max}_{\mathbf{F}}  \mathcal{F}_R\left(\mathbf{F}|\mathbf{w}^{(t+1)},\mathbf{W}_{RF}^{(t+1)},\mathbf{u}_r^{(t)} \right),  s.t.  \mathcal{F}_k\left(\mathbf{F}|u_k^{(t)} \right)\geq \zeta^{(t)}, k\in [1,K], 
				(\ref{cont05}),\label{probFt1}\\
				&\mathbf{u}_r^{(t+1)}
				=\frac{ \sigma_t\mathbf{F}^{(t+1),\HH}\mathbf{A}_{R}^{\HH}\mathbf{W}_{RF}^{(t+1)} \mathbf{w}^{(t+1)}  }{\mathcal{P}_{Q}\left(\mathbf{w}^{(t+1)},\mathbf{W}_{RF}^{(t+1)},\mathbf{F}^{(t+1)} \right)},\label{urt1}\\
				&u_{k}^{(t+1)}
				=\frac{\mathbf{f}_c^{(t+1),\HH}\mathbf{h}_{c,k}}{\left\Vert \mathbf{h}_{c,k}^\HH \mathbf{F}_R^{(t+1)}  \right\Vert^2+\sum_{i\neq k}\left|\mathbf{h}_{c,k}^\HH\mathbf{f}_{c,i}^{(t+1)}  \right|^2
					+\sigma_c^2},\\
				&\zeta^{(t+1)}=\min_{k \in [1,K]} \kappa_c\gamma_k (\mathbf{F}^{(t+1)}).\label{zetat1}
			\end{align}
		\end{subequations}
	
	\textbf{Remark 1} (\emph{Convergence Analysis}): We have the following proposition regarding the convergence of the proposed algorithm. 
	\begin{proposition}
		\label{PROconverge}
		The iteration in (\ref{IterationCore}) creates a non-decreasing sequence 
		\begin{equation}
			\begin{split}
				\mathcal{L}^{(t)}
				&=\mathcal{L}\left(\mathbf{w}^{(t)},\mathbf{W}_{RF}^{(t)},\mathbf{F}^{(t)} \right)=\mathcal{F}\left(\mathbf{w}^{(t)},\mathbf{W}_{RF}^{(t)},\mathbf{F}^{(t)},\mathbf{u}_r^{(t)},u_{k}^{(t)} \right),\\
			\end{split}
		\end{equation}  
		which converges to the stationary point of (\ref{Problem0}), i.e., $\mathcal{L}	^{\star}=\mathcal{L}\left(\mathbf{w}^{\star},\mathbf{W}_{RF}^{\star},\mathbf{F}^{\star} \right)$. 
	\end{proposition}
	
	\textit{Proof}: See Appendix \ref{IterConv}. \hfill $\blacksquare$

	The convergence and optimality of the proposed algorithm  are guaranteed by Proposition 1 and  \cite[Theorem 4]{8314727}, respectively.
	The next issue is to solve (\ref{probwt1})-(\ref{probFt1}).
	\subsection{Update $\mathbf{w}^{(t+1)}$}
	First, we reformulate (\ref{lossfunc})  w.r.t. $\mathbf{w}$ as
	\begin{equation}
		\label{16}
		\begin{split}
			&\mathcal{F}_R\left(\mathbf{w}|\mathbf{W}_{RF},\mathbf{F},\mathbf{u}_r \right)=2 \kappa_r \Re\left(   \mathbf{w}^\HH \mathbf{a}_{w} \right)- \kappa_r\Vert\mathbf{u}_r\Vert^2  
			\mathbf{w}^\HH \mathbf{B}_{w}  \mathbf{w}+const,\\
		\end{split}
	\end{equation}
	where
	\begin{equation}
		\begin{split}
			&\mathbf{a}_w= \sigma_t \mathbf{W}_{RF}^\HH\mathbf{A}_{R}\mathbf{F}  \mathbf{u}_r, \quad\mathbf{B}_{w}=\sum\limits_{l=1}^{L}\sigma_{c,l}^2  \mathbf{W}_{RF}^\HH\mathbf{A}_{c,l}\mathbf{F}\mathbf{F}^\HH\mathbf{A}_{c,l}^\HH \mathbf{W}_{RF}+\sigma_n^2 \mathbf{W}_{RF}^\HH  \mathbf{W}_{RF} . 
		\end{split}
	\end{equation}
	Then the problem in (\ref{probfracF}) is reformulated as the maximization of (\ref{16}).
	This problem can be efficiently solved by the Lagrange multiplier method.
	Specifically, we introduce a penalty function to reformulate the problem in (\ref{16}) as an unconstrained optimization problem that minimizes
	\begin{equation}
		\begin{split}
			L_{w}^{(t)}(\mathbf{w})=&-\mathcal{F}_R\left(\mathbf{w}|\mathbf{W}_{RF}^{(t)},\mathbf{F}^{(t)},\mathbf{u}_r^{(t)} \right)+ \gamma_{w} \left(\mathbf{w}^\HH\mathbf{w} -1 \right),\\
		\end{split}
		\label{objsubiF3}
	\end{equation}
	where $\gamma_{w} \geq 0$ is the Lagrange penalty coefficient.	Note that (\ref{objsubiF3}) is convex w.r.t. $\mathbf{w}$. The minimizer of (\ref{objsubiF3}) can be obtained by solving $\nabla_{\mathbf{w}} L_{w}^{(t)}(\mathbf{w})=0$, i.e.,
	\begin{equation}
		\begin{split}
			\mathbf{w}^{(t,\star)}=\frac{1}{\Vert\mathbf{u}_r^{(t)}\Vert^2 }\left(\mathbf{B}_{w}^{(t)}+\gamma_w\mathbf{I}\right)^{-1}\mathbf{a}_w^{(t)}.
		\end{split}
		\label{wstar}
	\end{equation}
	
	Note that $\mathbf{w}^{(t,\star)}$ depends on  $\gamma_{w}$. Therefore, in the rest of this subsection, we focus on determining $\gamma_{w}$. By performing the eigen-decomposition $ \mathbf{B}_w^{(t)}=\mathbf{V}_w\mathbf{\Lambda}_w\mathbf{V}_w^\HH$ and based on the complementary Karush–Kuhn–Tucker (KKT) condition, we have
	\begin{equation}
		\begin{split}
			\mathbf{w}^{(t,\star),\HH}\mathbf{w}^{(t,\star)}&= \mathbf{a}_w^{(t),\HH} \mathbf{V}_w\left(\mathbf{\Lambda}_w+\lambda_{w,1}\mathbf{I} \right)^{-2}\mathbf{V}_w^\HH \mathbf{a}_w^{(t)}=\sum_{i=1}^{N_{RF}^r} \frac{\left|\mathbf{v}_{w,i}^\HH \mathbf{a}_w^{(t)}\right|^2}{\left(\lambda_{w,i}+\gamma_{w}\right)^2} = 1,
		\end{split}	
		\label{lambdaF1}
	\end{equation}
	where $\mathbf{v}_{w,i}$ denotes the $i$th column of $\mathbf{V}_{w}$ and $\lambda_{w,i}$ is the $(i,i)$th entry of $\mathbf{\Lambda}_w$. It is easy to check that $\tr\left( \mathbf{w}^{(t+1)}\mathbf{w}^{(t+1),\HH}  \right)$ is monotonic w.r.t. $\gamma_{w}$.
	We thus utilize the bisection method to find a suitable $\gamma_{w}$ to make $||\mathbf{w}^{{(t,\star)}}||=1$. We then update $\mathbf{w}^{(t+1)}=\mathbf{w}^{{(t,\star)}}$.
	
	\subsection{Update $\mathbf{W}_{RF}^{(t+1)}$}
	\label{updatewRF}
	Note that $\mathcal{P}_Q$ is dependent on $\mathbf{W}_{RF}$. Therefore, we first simplify the formulation in (\ref{lossfunc}). Defining the vectorization of $\mathbf{W}_{RF}$ as $\mathbf{w}_{RF}=\mathrm{vec}(\mathbf{W}_{RF})$, the objective function in (\ref{lossfunc})  w.r.t. $\mathbf{w}_{RF}$ is then reformulated as
	\begin{equation}
		\begin{split}
			&\mathcal{F}\left(\mathbf{w}_{RF}|\mathbf{F},\mathbf{u}_r,u_{k} \right)
			=2 \kappa_r \Re\left(  \mathbf{w}_{RF}^\HH \mathbf{a}_{w_{RF}} \right) -\kappa_r\Vert\mathbf{u}_r\Vert^2  
			\mathbf{w}_{RF}^\HH \mathbf{B}_{w_{RF}}  \mathbf{w}_{RF}+const,\\
		\end{split}
		\label{fwRF}
	\end{equation}
	where
	\begin{equation}
		\begin{split}
			&\mathbf{a}_{w_{RF}}=\sigma_t \mathrm{vec} \left(\mathbf{A}_{R}\mathbf{F} \mathbf{u}_r\mathbf{w}^\HH\right),\quad\mathbf{B}_{w_{RF}}= \mathbf{w}^* \mathbf{w}^\TT \otimes \left(\sum\limits_{l=1}^{L} \sigma_{c,l}^2 \mathbf{A}_{c,l}\mathbf{F} \mathbf{F}^\HH\mathbf{A}_{c,l}^\HH+\sigma_n^2 \mathbf{I}\right).
		\end{split}
	\end{equation}
	From (\ref{probWRFt1}), we can observe that the objective function is convex w.r.t. $\mathbf{w}_{RF}$, whereas the feasible set $\mathcal{M}^{N_{r}N_{RF}^r\times 1}$ is still non-convex. Fortunately, $\mathcal{M}^{N_{r}N_{RF}^r\times 1}$ is known as the complex circle manifold (CCM) so that (\ref{probWRFt1}) can be addressed by the manifold-based method.
	
	The manifold-based method updates the variable within the tangent space $\mathcal{T} \mathcal{M}$. By updating along the tangent space with a small enough step, the new point is almost within $\mathcal{M}$.
	For the manifold in (\ref{cont06}), its corresponding tangent space is given as
	\begin{equation}
		\begin{split}
			&\mathcal{T}_{\mathbf{w}_{RF}} \mathcal{M}^{N_{r}N_{RF}^r\times 1}=\{\mathbf{x} \in \mathbb{C}^{N_{r}N_{RF}^r\times 1}| \Re (\mathbf{x} \circ \mathbf{w}_{RF})=\mathbf{0} \}.
		\end{split}
	\end{equation}
	Resembling the gradient-based method, the manifold-based method will find a direction from the tangent space where the objective function decreases most steeply (for minimization problems), i.e., the negative Riemannian gradient direction. For the manifold $\mathcal{M}^{N_{r}N_{RF}^r\times 1}$, the Riemannian gradient at $\mathbf{w}_{RF}$ is a tangent vector  given as the orthogonal projection of the Euclidean gradient $\nabla \mathcal{G}_{w}^{(t)}(\mathbf{w}_{RF})$ onto the tangent space \cite{absil2009optimization}, i.e.,
	\begin{equation}
		\begin{split}
			&\mathbf{\eta}(\mathbf{w}_{RF})=\mathrm{grad} \mathcal{G}_{w}^{(t)}(\mathbf{w}_{RF})=\nabla \mathcal{G}_{w}^{(t)}(\mathbf{w}_{RF})- \Re(\nabla \mathcal{G}_{w}^{(t)}(\mathbf{w}_{RF}) \circ \mathbf{w}_{RF} ) \circ
			\mathbf{w}_{RF},
		\end{split}
		\label{RiemGrad}
	\end{equation}
	where 	$\nabla \mathcal{G}_{w}^{(t)}(\mathbf{w}_{RF})=-2\kappa_r\mathbf{a}_{w_{RF}}^{(t)}+ 2\kappa_r \Vert\mathbf{u}_r^{(t)}\Vert^2 \mathbf{B}_{w_{RF}}^{(t)} \mathbf{w}_{RF}$
	denotes the Euclidean gradient.
	The Riemannian gradient in the tangent space is the optimization direction that shifts the manifold the least. 
	In practice, a retraction is needed to remap the updated points from the tangent space onto the manifold. 
	The retraction of a tangent vector $\beta \mathbf{d} \in \mathcal{T}_{\mathbf{w}_{RF}}\mathcal{M}^{N_{r}N_{RF}^r\times 1}$ at $\mathbf{w}_{RF}$ is
	\begin{equation}
		\begin{split}
			\mathcal{P}: &\mathcal{T}_{\mathbf{w}_{RF}}\mathcal{M}^{N_{r}N_{RF}^r\times 1}\to \mathcal{M}^{N_{r}N_{RF}^r\times 1}\\
			& \beta \mathbf{d} \to \left(\mathbf{w}_{RF}+\beta \mathbf{d} \right) \circ \frac{\mathbf{1}}{|\mathbf{w}_{RF}+\beta \mathbf{d}|_e},
		\end{split}
		\label{funcP}
	\end{equation}
	where $\frac{\mathbf{1}}{|\mathbf{x}|_e}\in \mathcal{R}^{N_{r}N_{RF}^r\times 1}$ denotes a vector whose $i$th entry is ${1}/{|x_{i}|}$, and $\beta$ represents the Armijo step \cite[Definition 4.2.2]{absil2009optimization}.
	
	To solve (\ref{probWRFt1}), we propose the manifold-based method summarized in Algorithm \ref{ALG1}, whose convergence is guaranteed by \cite[Theorem 4.3.1]{absil2009optimization}. Algorithm \ref{ALG1} provides the update $\mathbf{w}_{RF}^{(t+1)}=\mathbf{w}_{RF}^{{(t,\star)}}$.
	
	\begin{algorithm}[t] 
		\caption{Proposed Manifold-based Method to obtain $\mathbf{w}_{RF}^{{(t,\star)}}$} 
		\textbf{Input:} An initial point $\mathbf{w}_{RF}^{(t,0)}=\mathbf{w}_{RF}^{(t)}$ and $\mathbf{d}^{(0)}=-\mathbf{\eta} \left(\mathbf{w}_{RF}^{(t,0)} \right)$.
		
		\textbf{Repeat}
		\begin{enumerate} 
			\item Compute $\beta^{(m)}$ via the Armijo line search step \cite[Definition 4.2.2]{absil2009optimization}.
			\item Update $\mathbf{w}_{RF}^{(t,m+1)}=\mathcal{P}(\beta^{(m)} \mathbf{d}^{(m)})$ via (\ref{funcP}).
			\item Compute the Riemannian gradient $\mathbf{\eta} \left(\mathbf{w}_{RF}^{(t,m+1)} \right)$ via (\ref{RiemGrad}).
			\item Update the optimization direction $\mathbf{d}^{(m+1)}=-\mathbf{\eta} \left(\mathbf{w}_{RF}^{(t,m+1)} \right)$.
			\item $m \gets m+1$.
		\end{enumerate} 
		\textbf{Until} Convergence criterion is met.\\
		\textbf{Output:} The optimal solution $\mathbf{w}_{RF}^{{(t,\star)}}$.
		\label{ALG1}
	\end{algorithm}

	\subsection{Update $\mathbf{F}^{(t+1)}$} 
	We first reformulate the problem w.r.t. $\mathbf{F}$ as a quadratically constrained quadratic programming (QCQP) which is a subclass of semi-definite programming (SDP) \cite{vandenberghe1996semidefinite}.
	Define the vectorization of $\mathbf{F}$ as $\mathbf{f}=\mathrm{vec}\left(\mathbf{F}\right)$.
	Omitting some constants,  (\ref{lossfunc}) w.r.t. $\mathbf{f}$ is  rewritten as
	\begin{equation}
		\begin{split}
			&\mathcal{F}_R\left(\mathbf{f}|\mathbf{w},\mathbf{W}_{RF}, \mathbf{u}_r \right)=2\kappa_r\Re(\mathbf{a}_F^\HH  \mathbf{f})-\kappa_r\Vert\mathbf{u}_r\Vert^2  \mathbf{f}^\HH  \mathbf{B}_F \mathbf{f}+const, 
		\end{split}
		\label{fRrefo}
	\end{equation}
	where
	\begin{equation}
		\begin{split}
			&\mathbf{a}_F= \sigma_t \mathrm{vec}\left(\mathbf{A}_{R}^\HH \mathbf{W}_{RF} \mathbf{w}\mathbf{u}_r^\HH \right),\quad \mathbf{B}_F=\mathbf{I} \otimes \left(\sum\limits_{l=1}^{L} \sigma_{c,l}^2\mathbf{A}_{c,l}^\HH\mathbf{W}_{RF}\mathbf{w}\mathbf{w}^\HH \mathbf{W}_{RF}^\HH\mathbf{A}_{c,l}\right).
		\end{split}
	\end{equation}
	Then (\ref{lossfuncCk}) can be rewritten as
	\begin{equation}
		\begin{split}
			&\mathcal{F}_k \left(\mathbf{f} |u_{k} \right)=2\kappa_c\Re\left(  \mathbf{a}_{F_k}^\HH \mathbf{f} \right)-\kappa_c \mathbf{f}^\HH \mathbf{B}_{F_k} \mathbf{f} -\kappa_c|u_k|^2 \sigma_c^2,
		\end{split}
		\label{fkrefo}
	\end{equation}
	where 
	\begin{equation}
		\begin{split}
			&\mathbf{a}_{F_k}=u_{k}^*\mathbf{e}_k\otimes \mathbf{h}_{c,k},\quad\mathbf{B}_{F_k}=||u_{k}||^2 (\mathbf{1}-\mathbf{e}_k )(\mathbf{1}-\mathbf{e}_k )^\TT  \otimes \mathbf{h}_{c,k}\mathbf{h}_{c,k}^\HH,
		\end{split}
	\end{equation}
	with $\mathbf{e}_k=[\underbrace{0,\cdots,0}_{k-1},1,\underbrace{0,\cdots,0}_{N_s-k}]^\TT.$

	By substituting (\ref{fRrefo}) and (\ref{fkrefo}) into (\ref{probFt1}), the resultant problem is a typical QCQP because the objective function and all constraints are reformulated as a linear or quadratic form. 
	This problem can be easily solved by the well-known CVX toolbox \cite{cvx}. We then update $\mathbf{F}^{(t+1)}$ by rearranging $\mathbf{f}^{{(t,\star)}}$.
	
	\textbf{Remark 2}:		
	The proposed AO-based method consists of six sub-optimization problems, i.e., (\ref{probwt1})-(\ref{zetat1}), where the computational cost is dominated by (\ref{probwt1}), (\ref{probWRFt1}) and (\ref{probFt1}). The first step aims to obtain the digital baseband processor $\mathbf{w}$, whose main computational cost comes from the inverse operation  of the $N_{RF}^r \times N_{RF}^r$ matrix. Therefore, the computational cost of the first step is about $\mathcal{O}((N_{RF}^r)^3 I_1)$ where $I_1$ denotes the number of iterations to update $\mathbf{w}$. The second step is a manifold-based optimization to obtain $\mathbf{W}_{RF}$, whose computational cost is about $\mathcal{O}(N_r N_{RF}^r I_{2})$ where $I_2$ denotes the number of iterations required by the bisection method to update $\mathbf{W}_{RF}$  \cite{9234098}. The third step is a QCQP solved by the CVX toolbox, whose computational cost is about $\mathcal{O}((N_t N_s)^3 I_{3})$, where $I_3$ denotes the required number of iterations to update $\mathbf{F}$. In this case, the total computational cost of the proposed AO-based method is about $\mathcal{O}(((N_{RF}^r)^3 I_1+N_r N_{RF}^r I_{2}+(N_t N_s)^3 I_{3})T)$, where $T$ denotes the number of iterations required by the proposed optimization problem in (\ref{probfracF}). It is observed from experiments that $T$ is usually no more than 10.
	
	Meanwhile, the performance of the AO-based  method depends on the weighting coefficients $\kappa_c$ and $\kappa_R$, but it is hard to adjust them to meet a desired communication or sensing performance. These issues motivate us to find other methods. One option is to use other optimization methods, such as the alternating direction method of multipliers (ADMM). However, such methods also involve an internal iteration process whose computational cost is still high since the dimension of $\mathbf{f}$ is large. Moreover, their performance also depends on some intermediate parameters. In the next section,  we derive some linear transceiver structures to reduce the complexity and reveal more physical sights. 
	
	\section{Linear Transceiver Design}
	In this section, we derive two sub-optimal transceiver structures to reduce the computational cost. These methods aim to maximize the sensing performance with given communication requirements. 
	
	\subsection{Linear Transmitter Design}
	\subsubsection{ZF Transmitter}
	ZF is a well-established beamforming method. In the concerned PMN-TMT, the channel matrix for the UEs is given as $\mathbf{H}_c
	=\left[ \mathbf{h}_{c,1},\cdots,\mathbf{h}_{c,K} \right] \in \mathbb{C}^{N_t\times K}.$
	Then, the zero forcing precoder can be written as
	\begin{equation}
		\label{Fzfdef}
		\begin{split}
			\mathbf{F}_{\text{ZF}}=\mu \mathbf{H}_c(\mathbf{H}_c^\HH \mathbf{H}_c)^{-1}\in \mathbb{C}^{N_t\times K},
		\end{split}
	\end{equation}
	where 
	\begin{equation}
		\begin{split}
			\mu=\sqrt{\frac{P}{\mathrm{tr}\left[(\mathbf{H}_c^\HH \mathbf{H}_c)^{-1}\right]}}
		\end{split}
	\end{equation}
	denotes the normalized coefficient that guarantees $||\mathbf{F}_{\text{ZF}}||^2=P$.
	It can be validated that 	
	\begin{equation}
		\begin{split}
			\mathbf{h}_{c,i}^\HH \mathbf{f}_{\text{ZF},j}=\left\{\begin{matrix}
				\mu,i= j\\
				0,i\neq j\\
			\end{matrix}
			\right. , 
		\end{split}
	\end{equation}
	where $\mathbf{f}_{\text{ZF},j}$ denotes the $j$th column of $\mathbf{F}_{\text{ZF}}$, indicating that the  beam for one UE will not generate interference to others. 
	
	For an ISAC system, the channel between the BS and the ST is unknown. Thus, we leverage the steering vector towards the ST. Construct the ISAC `channel' matrix as
	\begin{equation}
		\label{Heisac}
		\begin{split}
			\mathbf{H}_e(\lambda_{a})
			=\left[ \mathbf{H}_{c},\lambda_{a}\mathbf{a}_T(\phi_t) \right] \in \mathbb{C}^{N_t\times (K+1)},
		\end{split}
	\end{equation}
	where $\lambda_{a}$ denotes a normalized coefficient to balance the amplitude of $\mathbf{a}_T(\phi_t)$.
	Then the ZF-ISAC precoder is given as
	\begin{equation}
		\begin{split}
			\mathbf{F}_{\text{ZF-ISAC}}=\mu_{a}(\lambda_{a}) \mathbf{H}_e(\lambda_{a})\left(\mathbf{H}_e^\HH(\lambda_{a}) \mathbf{H}_e(\lambda_{a})\right)^{-1},
		\end{split}
		\label{Fzfisac}
	\end{equation}
	where
	\begin{equation}
		\label{mudef}
		\begin{split}
			\mu_{a}(\lambda_{a})=\sqrt{\frac{P}{\mathrm{tr}\left[(\mathbf{H}_e^\HH(\lambda_{a}) \mathbf{H}_e(\lambda_{a}))^{-1}\right]}}.
		\end{split}
	\end{equation}
	\begin{lemma}
		\label{mualemma}
		The normalized coefficient $\mu_a(\lambda_{a})$ can be simplified as
		\begin{equation}
			\label{mualammadef}
			\begin{split}
				\mu_{a}(\lambda_{a})=\sqrt{\frac{P}{C_a + \frac{1}{\lambda_a^2 C_b}}},
			\end{split}
		\end{equation}
		where 
		\begin{equation}\nonumber
			\begin{split}
				C_a&=\tr(\mathbf{H}_c^\HH \mathbf{H}_c)^{-1}+\frac{\mathbf{a}_T^\HH(\phi_t) \mathbf{H}_c(\mathbf{H}_c^\HH \mathbf{H}_c)^{-2}\mathbf{H}_c^\HH\mathbf{a}_T(\phi_t)}{1-\mathbf{a}_T^\HH(\phi_t) \mathbf{H}_c(\mathbf{H}_c^\HH \mathbf{H}_c)^{-1}\mathbf{H}_c^\HH\mathbf{a}_T(\phi_t)},\\
				C_b&=1-\mathbf{a}_T^\HH(\phi_t) \mathbf{H}_c(\mathbf{H}_c^\HH \mathbf{H}_c)^{-1}\mathbf{H}_c^\HH\mathbf{a}_T(\phi_t).
			\end{split}
		\end{equation}
	\end{lemma}
	\emph{Proof}: See Appendix \ref{muaproof}. \hfill $\blacksquare$
	
	It can be checked that
	\begin{equation}
		\begin{split}
			&\mathbf{h}_{c,i}^\HH \mathbf{f}_{\text{ZF-ISAC},j}=\left\{\begin{matrix}
				\mu_a,i= j\\
				0,i\neq j\\
			\end{matrix}
			\right. 
			,\quad\mathbf{a}_T^\HH(\phi_t)\mathbf{f}_{\text{ZF-ISAC},j}=\left\{\begin{matrix}
				\frac{\mu_a}{\lambda_{a}},j= K+1\\
				0,j \neq K+1\\
			\end{matrix}
			\right. ,
		\end{split}
	\end{equation}
	which indicates that there is no interference between sensing and  communication.
	By substituting (\ref{Fzfisac}) into (\ref{gammak}), we have the SINR for $K$ UEs as
	\begin{equation}
		\begin{split}
			\gamma_1=\cdots=\gamma_K=\frac{\mu_{a}^2(\lambda_{a})}{\sigma_c^2}.
		\end{split}
	\end{equation}
	To guarantee the minimum SINR among all UEs is greater than a given threshold, i.e.,
	\begin{equation}
		\begin{split}
			\min_{k \in [1,K]}\gamma_k = \frac{\mu_{a}^2(\lambda_{a})}{\sigma_c^2} \geq \Gamma,
		\end{split}
	\end{equation}
	we can obtain $\lambda_{a}=\sqrt{\frac{1}{C_b(\frac{P}{\Gamma \sigma_c^2}-C_a)}}$
	where $\Gamma$ denotes the given threshold. 
	The  transmit power on the direction of the ST is thus given by
	\begin{equation}
		\label{ptgtzf0}
		\begin{split}
			P_{ZF,tgt}
			&\triangleq \mathbb{E}\left(||a_T^\HH(\phi_t) \mathbf{F}_{\text{ZF-ISAC}}\mathbf{s}||^2\right)= \frac{\mu_{a}^2(\lambda_{a})}{\lambda_{a}^2}=(P-\Gamma\sigma_c^2 C_a)C_b.
		\end{split}
	\end{equation}
	
	\subsubsection{Beam Synthesis Transmitter}
	Note that the communication signal can also be leveraged for sensing since the transmitted communication waveform is known by the TMT. Thus, although the sensing signals are not supposed to create interference to the UEs, the communication signals can be leaked to the direction of the ST. 
	In the following, we will investigate how the above observation can be exploited by the B-syn method \cite{9648341}.
	In particular, by the array response control, the beam pattern is synthesized to force the orientation and nulls of beams at the directions of the target
	and interference, respectively.
	For the given ISAC system, the B-syn precoder can be constructed as
	\begin{equation}
		\begin{split}
			&\mathbf{F}_{\text{B-syn}}\triangleq\left[\underbrace{\mathbf{f}_{\text{B-syn},c,1},\cdots,\mathbf{f}_{\text{B-syn},c,K}}_{\text{Communication}},\underbrace{\mathbf{f}_{\text{B-syn},R,1},\cdots,\mathbf{f}_{\text{B-syn},R,N_s-K}}_{\text{Sensing}}\right],
		\end{split}
		\label{Fbs}
	\end{equation}
	where
	\begin{equation}
		\begin{split}
			&\mathbf{f}_{\text{B-syn},c,i}=\alpha_i\mathbf{f}_{\text{ZF},i}+\beta_{i}\mathbf{f}_\perp, i=1,\cdots,K,\\
			&\mathbf{f}_{\text{B-syn},R,j}=\nu_j \mathbf{f}_\perp, j=1,\cdots,N_s-K,\\
		\end{split}
		\label{Fbt}
	\end{equation}
	denote the beamformers for the $K$ UEs and the radar target, respectively. Here $\mathbf{f}_{\text{ZF},i}$ denotes the $i$th column of $\mathbf{F}_{\text{ZF}}$ defined in (\ref{Fzfdef}). Note that the beamformer to the UE includes two parts, i.e., the ZF precoder to the $i$th UE ($\mathbf{f}_{\text{ZF},i}$) and the leaked communication signal towards the ST ($\mathbf{f}_\perp$). Here, $\alpha_i$, $\beta_i$ and $\nu_j$ denote the weighting coefficients for the communication terms, the leaked communication terms, and the dedicated sensing terms, respectively. To avoid interference from sensing to communication, the term $\mathbf{f}_\perp$ is expected to have the following property:
	\begin{equation}
		\begin{split}
			&\mathbf{H}_c^\HH\mathbf{f}_\perp=\mathbf{0},\quad\mathbf{a}_T^\HH(\phi_t)\mathbf{f}_\perp=1,
		\end{split}
		\label{ZFr}
	\end{equation}
	indicating that $\mathbf{f}_\perp$ should not impose interference to communication and should have a constant gain on the direction of the ST.
	We then choose $\mathbf{f}_\perp$ as the projection of $\mathbf{a}_T(\phi_t)$ in the null-space of $\mathbf{H}_c$, i.e.,
	\begin{equation}\label{Fperp}
		\begin{split}
			\mathbf{f}_\perp=\frac{\mathbf{a}_T(\phi_t)-\mathbf{H}_c(\mathbf{H}_c^\HH \mathbf{H}_c)^{-1}\mathbf{H}_c^\HH \mathbf{a}_T(\phi_t)}{1-\mathbf{a}_T^\HH(\phi_t)\mathbf{H}_c(\mathbf{H}_c^\HH \mathbf{H}_c)^{-1}\mathbf{H}_c^\HH \mathbf{a}_T(\phi_t)}.
		\end{split}
	\end{equation}
	It can be validated that $\mathbf{F}_{\text{ZF}}^\HH \mathbf{f}_{\perp} =\mathbf{0}$ and $||\mathbf{f}_\perp||^2=\frac{1}{C_b}$. From (\ref{Fbt}) and (\ref{ZFr}), we can observe the communication precoder $\mathbf{f}_{\text{B-syn},c,i}$ will transmit signals on the
	direction of the $i$th UE and the ST without interfering the other UEs. The sensing beamformer $\mathbf{f}_{\text{B-syn},R,j}$ will transmit signals on the direction of the ST without imposing interference to UEs. 
	
	The remaining issue is how to allocate power to these component beams. By substituting (\ref{Fbt}) into (\ref{gammak}), we have
	\begin{equation}
		\begin{split}
			\gamma_k=\frac{|\alpha_k|^2 \mu^2 }{\sigma_c^2}, k=1,\cdots,K.
		\end{split}
	\end{equation}
	It is desired to maximize the sensing performance with a minimal SINR for all UEs, i.e., $\min_{k \in [1,K]}\gamma_k \geq \Gamma$.
	Thus, we consider the equal-rate transmission, i.e., 
	\begin{equation}
		\begin{split}
			\alpha_1=\cdots=\alpha_K=\alpha_\star = \sqrt{\frac{\Gamma \sigma_c^2}{\mu^2}}.
		\end{split}
	\end{equation}
	The remaining task is to determine $\beta_{i}$ and $\nu_j$. We give the following proposition for allocating power to maximize the transmitted power on the direction of ST. 
	
	\begin{proposition}	\label{Ptgt}
		The optimal allocation is given as  $\nu_{j}=0,j=1,\cdots,N_s-K$ and
		\begin{equation}
			\begin{split}
				\beta_i=\sqrt{(P-\Gamma \sigma_c^2 \tr(\mathbf{H}_c^\HH \mathbf{H}_c)^{-1})C_b}\cdot\frac{\mathbf{f}_{\text{ZF},i}^\HH\mathbf{a}_T(\phi_t)}{||\mathbf{F}_{\text{ZF}}\mathbf{a}_T(\phi_t)||}, i=1,\cdots,K.
			\end{split}
		\end{equation}
	\end{proposition}
	The resultant transmit power on the direction of ST is given as
	\begin{equation}
		\begin{split}
			&P_{\text{B-syn},tgt}
			=2\alpha_\star\sqrt{P_q}||\mathbf{a}_{tgt}|| + C_{tgt},\\
		\end{split}
		\label{ptgt1}
	\end{equation}
where 
	\begin{equation}\label{ctgt}
	\begin{split}
		C_{tgt}&=\Gamma \sigma_c^2 \mathbf{a}_T^\HH(\phi_t) \mathbf{H}_{c}  \left(\mathbf{H}_{c}^\HH\mathbf{H}_{c}\right)^{-2}\mathbf{H}_{c}^\HH \mathbf{a}_T(\phi_t) +(P-\Gamma \sigma_c^2 \tr(\mathbf{H}_c^\HH \mathbf{H}_c)^{-1})C_b.
	\end{split}
\end{equation}

	\emph{Proof}: See Appendix \ref{ptgtproof}.

	Proposition 2 indicates that instead of designing dedicated sensing streams, we should add the sensing streams to the communication streams. As a result, the transmit signal can be given by $\sum_{i=1}^{K}\alpha_\star \mathbf{a}_T^\HH(\phi_t)\mathbf{f}_{\text{ZF},i}s_{c,i}+\sum_{i=1}^{K}\beta_{i} s_{c,i}=\sum_{i=1}^{K}\left(\alpha_\star \mathbf{a}_T^\HH(\phi_t)\mathbf{f}_{\text{ZF},i}+\beta_{i}\right)s_{c,i}$. 
 	It can be observed from (\ref{ptgt1}) that there are three energy terms in $P_{\text{B-syn},tgt}$, including the two terms in $C_{tgt}$. On the other hand, if we use a dedicated sensing stream, the transmit signal is given by $\sum_{i=1}^{K}\alpha_\star \mathbf{a}_T^\HH(\phi_t)\mathbf{f}_{\text{ZF},i}s_{c,i} +\sum_{j=1}^{N_s-K}\nu_{j} s_{R,j}$ where the received power is equal to $C_{tgt}$. Thus, the advantage of the B-syn scheme comes from the cross-term $2\alpha_\star\sqrt{P_q}||\mathbf{a}_{tgt}||$. We may regard this cross-term as the reuse or leaking of communication energy for the sensing purpose.

	Comparing the transmit power towards the ST by ZF and B-syn, we have the following proposition regarding the improvement of B-syn over ZF.  
	\begin{proposition}
		The improvement of the transmit power towards the ST by B-syn over ZF can be obtained  
		from (\ref{ptgtzf0}) and (\ref{ptgt1}) as
		\begin{equation}
			\label{ineq}
			\begin{split}
				P_{\text{B-syn},tgt}-P_{ZF,tgt}
				&=\frac{2\Gamma \sigma_c^2 }{\mu^2} ||\mathbf{F}_{\text{ZF}}^\HH \mathbf{a}_T(\phi_t)||^2+2\sqrt{\frac{\Gamma \sigma_c^2 P_q}{\mu^2}}||\mathbf{F}_{\text{ZF}}^\HH \mathbf{a}_T(\phi_t)||\geq 0.
			\end{split}
		\end{equation}
	\end{proposition}
	
	\textbf{Remark 3}: The improvement by B-syn over ZF will be zero only when
	\begin{enumerate}
		\item $\Gamma = 0$, indicating that the required communications performance is zero. Then all power will be allocated to sensing. 
		\item $\mathbf{F}_{\text{ZF}}^\HH \mathbf{a}_T(\phi_t)=\mathbf{0}$. Recalling (\ref{Fzfdef}), it is equivalent to $\mathbf{H}_{c}^\HH \mathbf{a}_T(\phi_t)=\mathbf{0}$, implying that the ST already falls into the null space of $\mathbf{H}_{c}$.
	\end{enumerate}
	In the above cases, B-syn is equivalent to ZF. On the other hand, the performance gap between B-syn and ZF will become larger if  $||\mathbf{F}_{\text{ZF}}^\HH \mathbf{a}_T(\phi_t)||^2$ is larger. Note that $\mathbf{F}_{\text{ZF}}^\HH \mathbf{a}_T(\phi_t)$ denotes the power of $\mathbf{F}_{\text{ZF}}$ on the ST direction. In general, if the ST is closer to one UE, the correlation between them gets larger and it is more efficient to leak power from  that UE to the ST. 
	
	\textbf{Remark 4}: B-syn and ZF are the linear precoders, whose computational cost mainly comes from matrix operations. Specially, the computational cost of (\ref{Fzfdef}) and (\ref{Fperp}) are dominated by the inverse operation  of $\mathbf{H}_c^\mathrm{H} \mathbf{H}_c$ with complexity $\mathcal{O}(K^3+K^2N_t) $. Thus, the  computational cost of (\ref{Fzfdef}) and (\ref{Fperp}) are about $\mathcal{O}(K^3+2K^2N_t) $ and $\mathcal{O}(K^3+2 K^2 N_t+2 K N_t)$, respectively, whereas that of the AO for updating $\mathbf{F}$ is about $\mathcal{O}((N_t N_s)^{3}I_3)$. Note that we choose $N_s=K$ for B-syn and ZF.
	This indicates that the proposed B-syn and ZF methods can significantly reduce the computational cost. Meanwhile, B-syn and ZF can update the precoder with a given communication requirement. ZF transmits a dedicated sensing data stream to the ST  and there is no interference between the dedicated sensing signal and the communication signals.
	In contrast, B-syn leaks energy from the communication signal to ST. From (\ref{ineq}), we can observe  that, it is better to leak communication energy to the ST direction than designing a dedicated sensing signal. In particular, B-syn uses less data stream to achieve a better sensing performance than ZF.
	
	\subsection{Receiver Design}
	In the above, we give two methods to design the transmitter. With a given transmitter structure, to maximize the SCNR, the design of the receiver $\mathbf{w}_{\text{eff}}$ is given by 
	\begin{equation}
		\begin{split}
			\max_{\mathbf{w}_{\text{eff}}} \;\text{SCNR}(\mathbf{w}_{\text{eff}})=\frac{\sigma_t^2 P_{tgt} \left| \mathbf{w}_{\text{eff}}^\HH \mathbf{a}_R(\phi_t) \right|^2}{\mathbf{w}_{\text{eff}}^\HH \mathbf{R}_{CN} \mathbf{w}_{\text{eff}}},
		\end{split}
	\end{equation}
	where $P_{tgt}$ denotes the power transmitted to the ST and $\mathbf{R}_{CN}= \sum\limits_{l=1}^{L}\sigma_{c,l}^2 \mathbf{A}_{c,l}\mathbf{F}\mathbf{F}^\HH\mathbf{A}_{c,l}^\HH +\sigma_n^2\mathbf{I}.$
	This problem is the classic MVDR beamforming problem, whose solution is given by \cite{1449208}
	\begin{equation}\label{weff}
		\begin{split}
			\mathbf{w}_{\text{eff}}=\frac{\mathbf{R}_{CN}^{-1} \mathbf{a}_R(\phi_t)}{\mathbf{a}_R^\HH(\phi_t)\mathbf{R}_{CN}^{-1} \mathbf{a}_R(\phi_t)}.
		\end{split}
	\end{equation}
	Then we	can adopt the fast optimization method in \cite{8646553} to obtain $\mathbf{W}_{RF}$ and $\mathbf{w}$ from $\mathbf{w}_{\text{eff}}$. Assembling the MVDR receiver with B-syn and ZF yields two linear transceiver structures, i.e., `B-syn $+$ MVDR' and `ZF $+$ MVDR'.

	\textbf{Remark 5}: The computational cost of (\ref{weff}) and the method in \cite{8646553} are about $\mathcal{O}((N_{RF}^{r})^3)$ and $\mathcal{O}(N_{RF}^{r} N_s I_4)$, respectively,  where $I_4$ denotes the number of required iterations. Thus, the total computational cost of `B-syn $+$ MVDR' and `ZF $+$ MVDR' are about $\mathcal{O}((N_{RF}^{r})^3+N_{RF}^{r} N_s I_4+K^3+ K^2 N_t)$ and $\mathcal{O}((N_{RF}^{r})^3+N_{RF}^{r} N_s I_4+K^3+2 K^2 N_t+2 K N_t)$, respectively. 
		Compared with the computational cost of AO, i.e., $\mathcal{O}(((N_{RF}^r)^3 I_1+N_r N_{RF}^r I_{2}+(N_t N_s)^3 I_{3})T)$, the linear transceiver structures are more computationally efficient.

	\section{Simulation}
	In this section, we show the performance of the proposed PMN-TMT with different transceiver structures. In the simulation, we consider a mmWave system operating at a carrier frequency of 28GHz. We assume that the ISAC system serves 3 single antenna UEs and 1 ST. The BS employs a ULA with $N_t = 128$ antennas. Unless specified otherwise, we also set $N_r = 128$ and $N_{RF}=4$. In this paper, we fix the distance between the BS and TMT as 50m. The distances between the BS and UEs are set as a random variable uniformly distributed in the range $[19,21]$m. The AOD and AOA of the ST, UEs and clutter patches are set as random variable uniformly distributed in the range  $[-\frac{\pi}{2},\frac{\pi}{2}]$.
	The noise power at the UEs and TMT are $\sigma_c^2$ and $\sigma_n^2$, respectively. 
	We set $\sigma_c^2=\sigma_n^2=-90$dBm, $\sigma_t^2/\sigma_n^2=20$dB and $\frac{1}{L}\sum_{l=1}^{L}\sigma_{c,l}^2/\sigma_n^2=30$dB.
	Recalling 	(\ref{channelmodel}), the channel between the BS and $k$th UE is modeled as \cite{9234098}
	\begin{equation}
		\begin{split}
			\mathbf{h}_{c,k}=\sqrt{\frac{N_{t}}{N_{p}}} \sum_{i=1}^{N_{p}} \beta_{k,i}^{(t)} \mathbf{a}_{T} (\phi_{k,i}^{(t)}), 
		\end{split}
	\end{equation}
	where $\beta_{k,i}^{(t)}\sim \mathcal{CN}(0,10^{-0.1\kappa})$ denotes the complex gain of the LOS path and $\kappa$ is the path loss given as $\kappa=a+10b\log_{10}(d)+\epsilon$ 
	with $d$ denoting the distance between the BS and the $k$th UE and $\epsilon \sim \mathcal{CN}(0,\sigma_{\epsilon}^2)$ \cite{6834753}. Following \cite{6834753}, we set  $a=61.4$, $b=2$, $\sigma_{\epsilon}=5.8$dB. $\beta_{k,i}^{(t)}\sim \mathcal{CN}(0,10^{-0.1(\kappa+\mu)})$ denotes the complex gain of the NLOS path and $\mu$ is the Rician factor \cite{7503970}. We model the small-scale fading as Rician, where the Rician factors is set as 7dB for LOS and 0dB for NLOS. Here, we set $N_p=4$.

	For the manifold optimization, we set the maximum number of iterations as $200$. The tolerance for the norm of the gradient between two iterations is $10^{-4}$. To terminate the iteration, the tolerance for the objective function between two iterations is $10^{-2}$ and the maximum number of iterations is $20$.

	\subsection{System Performance}
		\begin{figure}[t]
		\centering
		\includegraphics[width=3.2in]{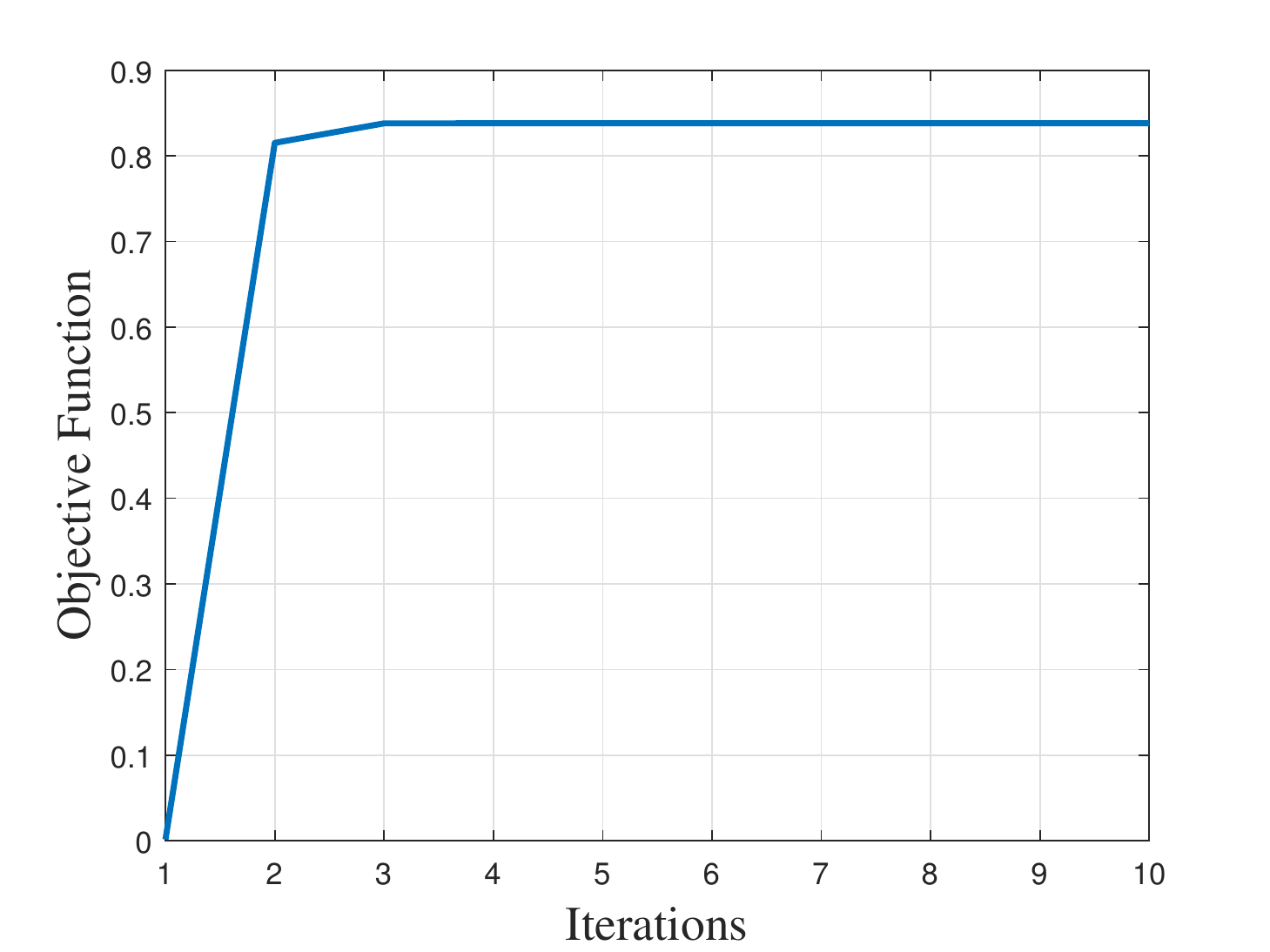}
		\caption{The objective function $\mathcal{L}(\mathbf{w},\mathbf{W}_{RF},\mathbf{F})$ over iterations.}
		\label{fig_Conv}
	\end{figure}
	
	In Sec. III, we prove the convergence of the proposed AO method. Here, we first show its convergence through the simulation. 
	Fig. \ref{fig_Conv} illustrates the convergence behavior of the average objective function  with  AO. Here we set $N_r=32$, $\kappa_c=0.5$ and 500 Monte-Carlo experiments are performed. 
	It can be observed that  the iteration will converge in about 4 rounds.

	\begin{figure}[t]
		\centering
		\includegraphics[width=3.2in]{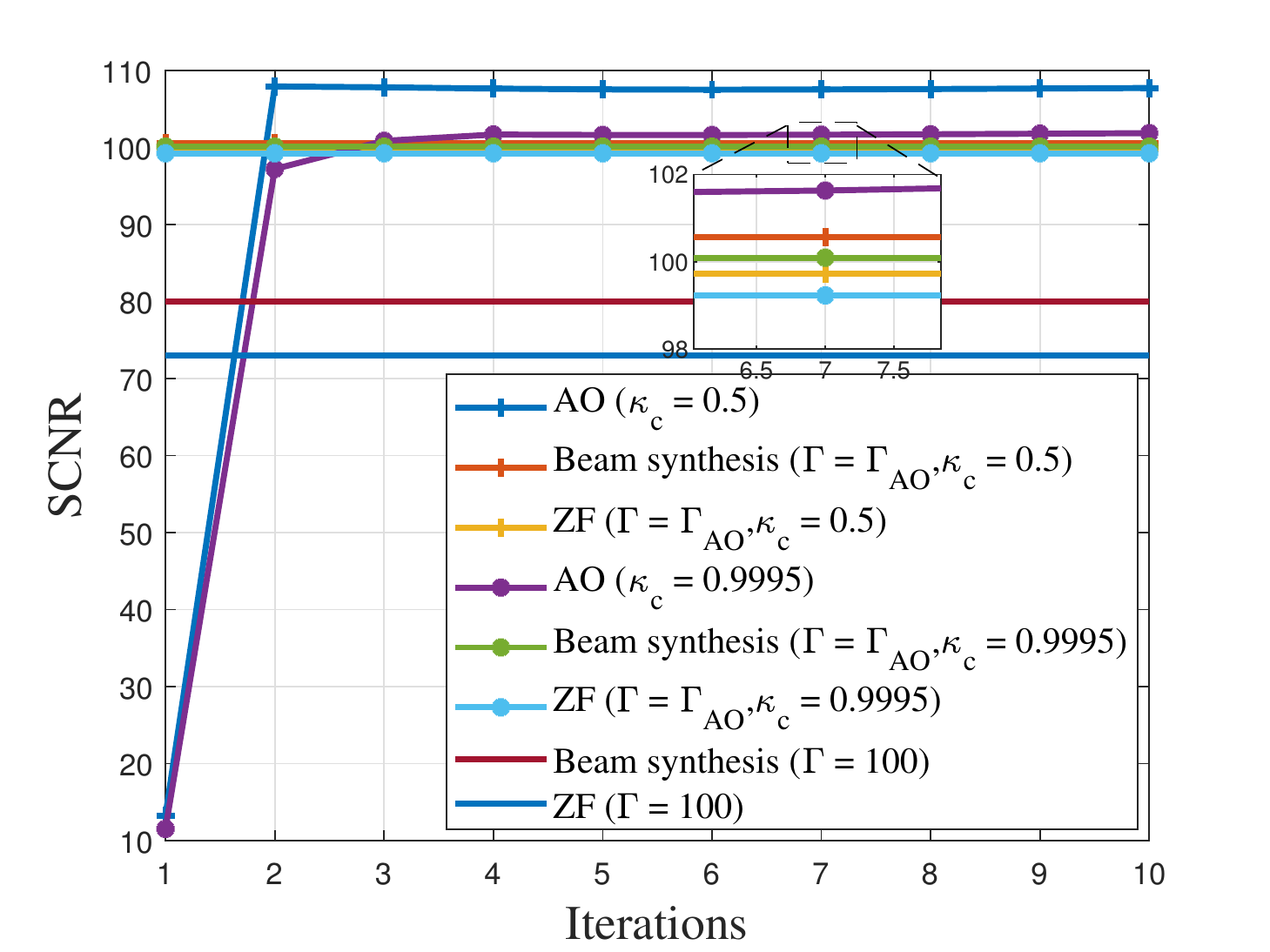}
		\caption{Comparison of different methods for updating $\mathbf{F}$.}
		\label{fig_BT}
	\end{figure}

	Fig. \ref{fig_BT} shows the average sensing performance of different transceiver structures with different levels of communication performance requirement with $N_r=32$. For each curve, 800 Monte-Carlo experiments are performed. 
	The legend `AO ($\kappa_c = C$)'  denotes the AO method proposed in Sec. III with $\kappa_c = C$. The legend `Beam synthesis' and `ZF' represent the proposed `B-syn + MVDR' and ‘ZF + MVDR’ transceivers with a given $\Gamma$, respectively. 
	In particular, $\Gamma=\Gamma_{\text{AO},\kappa_c=C}$ means that we fix $\Gamma$ to be the same as that of the `AO' with $\kappa_c=C$. Note that $\Gamma_{\text{AO},\kappa_c=0.5}<\Gamma_{\text{AO},\kappa_c=0.9995}<50$. We have several observations regarding the performance comparison.

	\textbf{AO vs. B-syn vs. ZF:} By comparing `AO ($\kappa_c=0.5$)', `Beam synthesis ($\Gamma=\Gamma_{\text{AO},\kappa_c=0.5}$)' and `ZF ($\Gamma=\Gamma_{\text{AO},\kappa_c=0.5}$)', we can observe that the performance of AO outperforms both B-syn and ZF, while B-syn is better than ZF. The same conclusion can be obtained by comparing `AO ($\kappa_c=0.9995$)', `Beam synthesis ($\Gamma=\Gamma_{\text{AO},\kappa_c=0.9995}$)' and `ZF ($\Gamma=\Gamma_{\text{AO},\kappa_c=0.9995}$)'. This is mainly due to different transceivers' tolerance for the interference between UEs and the ST. In particular, AO does not force the beams towards different UEs and the ST to be completely orthogonal. B-syn requires the orthogonality between UEs to completely eliminate the multi-UE interference but allows leakage from the communication signal to the ST. On the other hand, ZF strictly constrains the UEs and ST not to affect each other. In particular, the sensing performance will degrade when the orthogonality constraint is stronger.
	
	\textbf{AO vs. B-syn with Different Communication Requirements:} Comparing  `AO ($\kappa_c=0.5$)' and `Beam synthesis ($\Gamma=\Gamma_{\text{AO},\kappa_c=0.5}$)' with `AO ($\kappa_c=0.9995$)' and `Beam synthesis ($\Gamma=\Gamma_{\text{AO},\kappa_c=0.9995}$)', we can observe the gap between AO and B-syn will become smaller when $\kappa_c$ and the corresponding $\Gamma$ increase. This indicates that when the communication requirement is high, B-syn will behave similarly as AO. This is because when the communication requirement is low, the UEs can tolerate higher interference and thus it is not necessary to completely eliminate the multiuser interference. As the communication requirement increases, the multi-UE interference is more critical and forces AO to avoid it like B-syn. Thus, the gap is getting smaller. Under such circumstances, B-syn is preferable since it has much lower computational complexity than AO.

	\textbf{B-syn vs. ZF with Different Communication Requirements:} Comparing  `Beam synthesis ($\Gamma=\Gamma_{\text{AO},\kappa_c=0.5}$)' and `ZF ($\Gamma=\Gamma_{\text{AO},\kappa_c=0.5}$)' with `Beam synthesis ($\Gamma=50$)' and `ZF ($\Gamma=50$)', we can observe that, as $\Gamma$ increases, the gap between B-syn and ZF will become larger. Note here $\Gamma=50$ corresponds to a higher communication requirement than $\Gamma=\Gamma_{\text{AO},\kappa_c=0.5}$. This indicates that, compared with ZF, B-syn can achieve better sensing performance with the same communication requirement, and as the communication requirement $\Gamma$ increases,  the power improvement in (\ref{ineq}) becomes larger.

	\begin{figure}[h]
		\centering
		\includegraphics[width=3.2in]{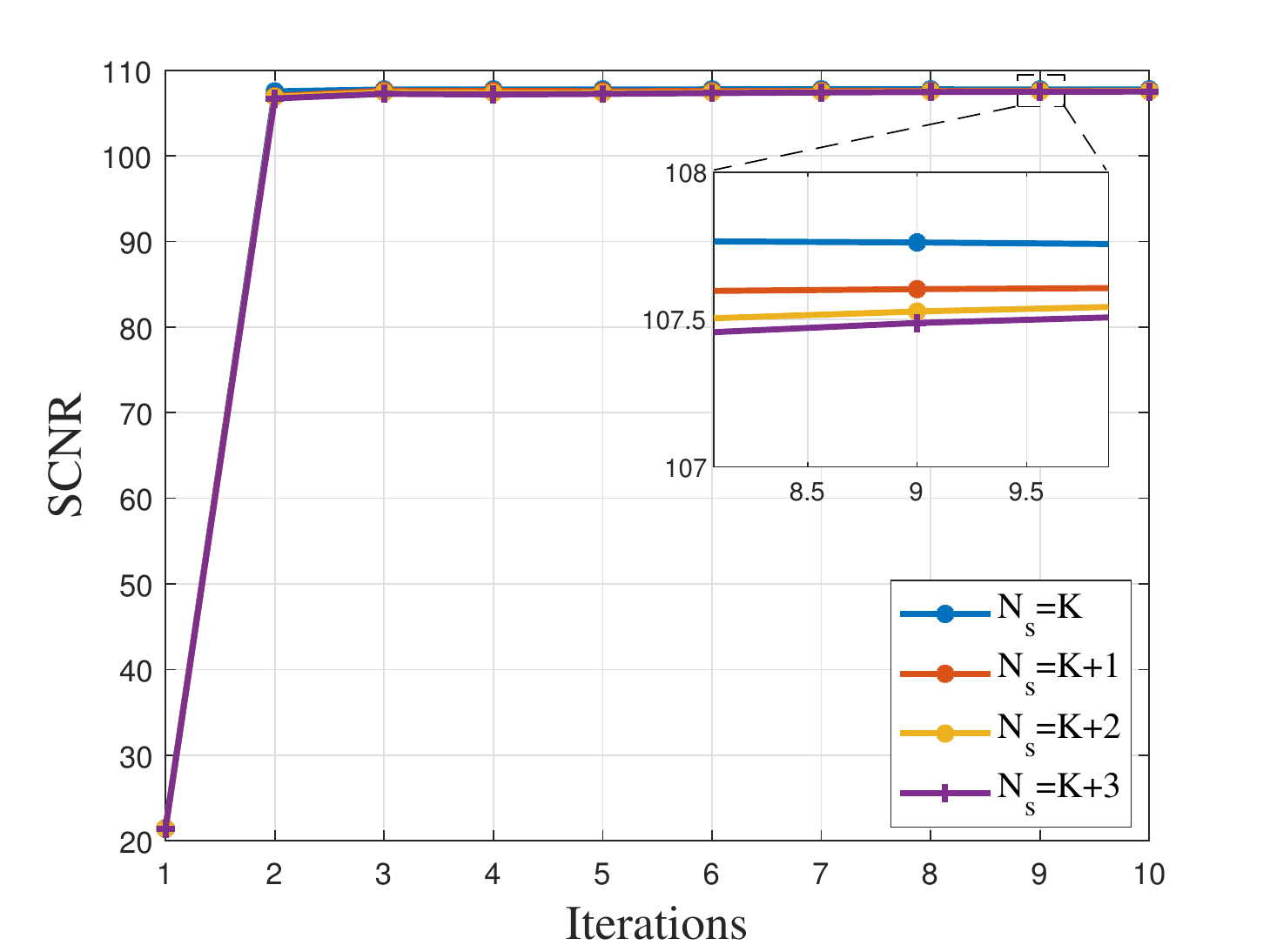}
		\caption{Effect of $N_s$ for AO.}
		\label{fig_Ns}
	\end{figure}

	\textbf{Energy Leaking:} In Sec. IV, we proved that it is more efficient to leak communication energy to the ST than sending a dedicated sensing signal, i.e., B-syn outperforms ZF. However, it is hard to prove this property for AO. In Fig. \ref{fig_Ns}, we show the sensing performance of AO with different number of data streams $N_s$ when $\kappa_c=0.5$. Note that the case $N_s=K$ indicates there are $K$ communication data streams and no dedicated sensing signal, while the case $N_s=K+C$ indicates that there are $C$ data streams for sensing. It can be observed that, as $N_s$ increases, the performance of AO becomes slightly worse, implying that the dedicated sensing signal is also less efficient for AO.
	
	\subsection{Physical Insights}
	In this section, we reveal some physical insights by looking into the beam patterns with different transceiver structures.

	\begin{figure*}[h]
		\centering
		\subfloat[]{\includegraphics[width=3.2in]{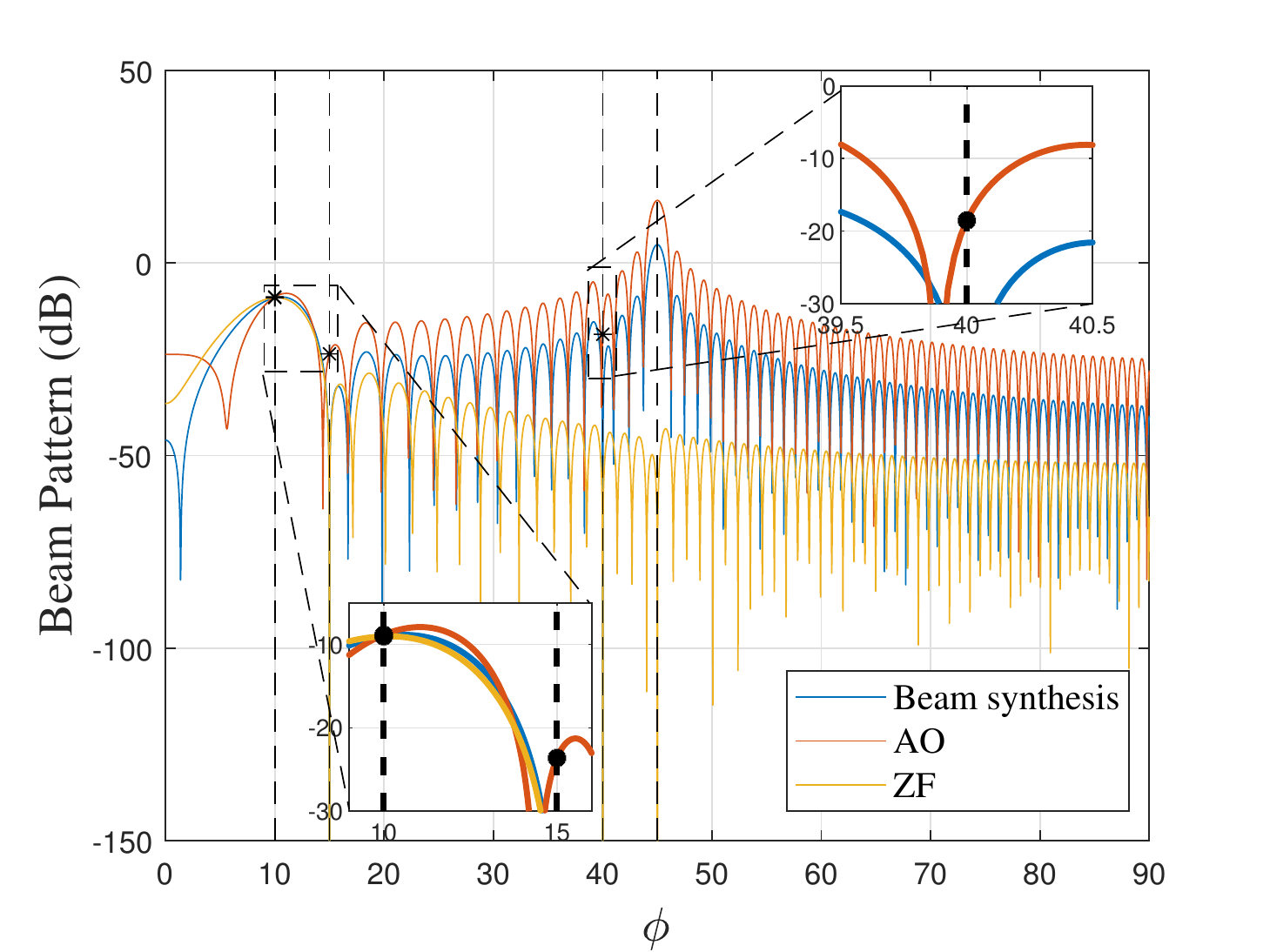}}\
		\subfloat[]{\includegraphics[width=3.2in]{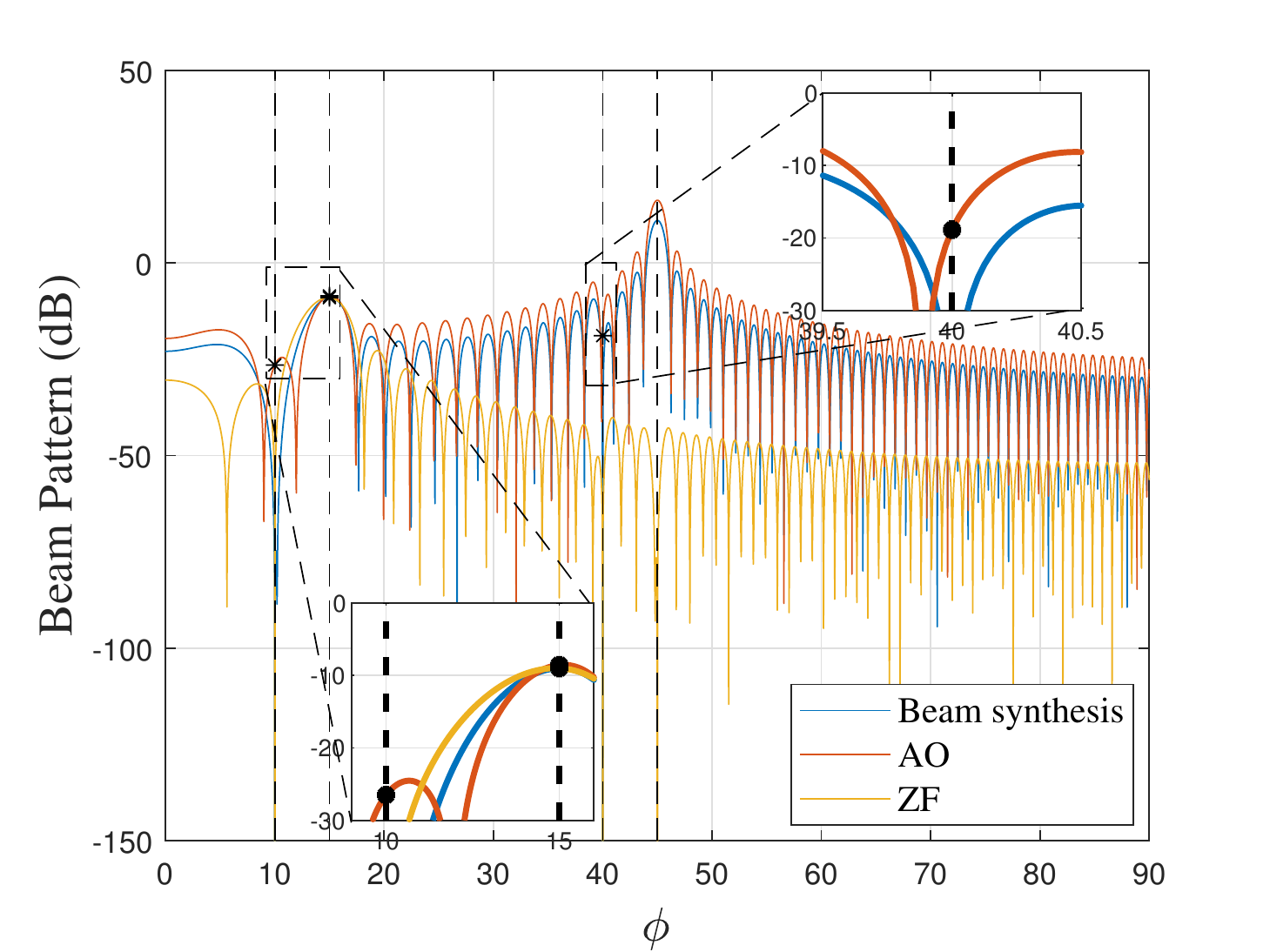}}\
		\subfloat[]{\includegraphics[width=3.2in]{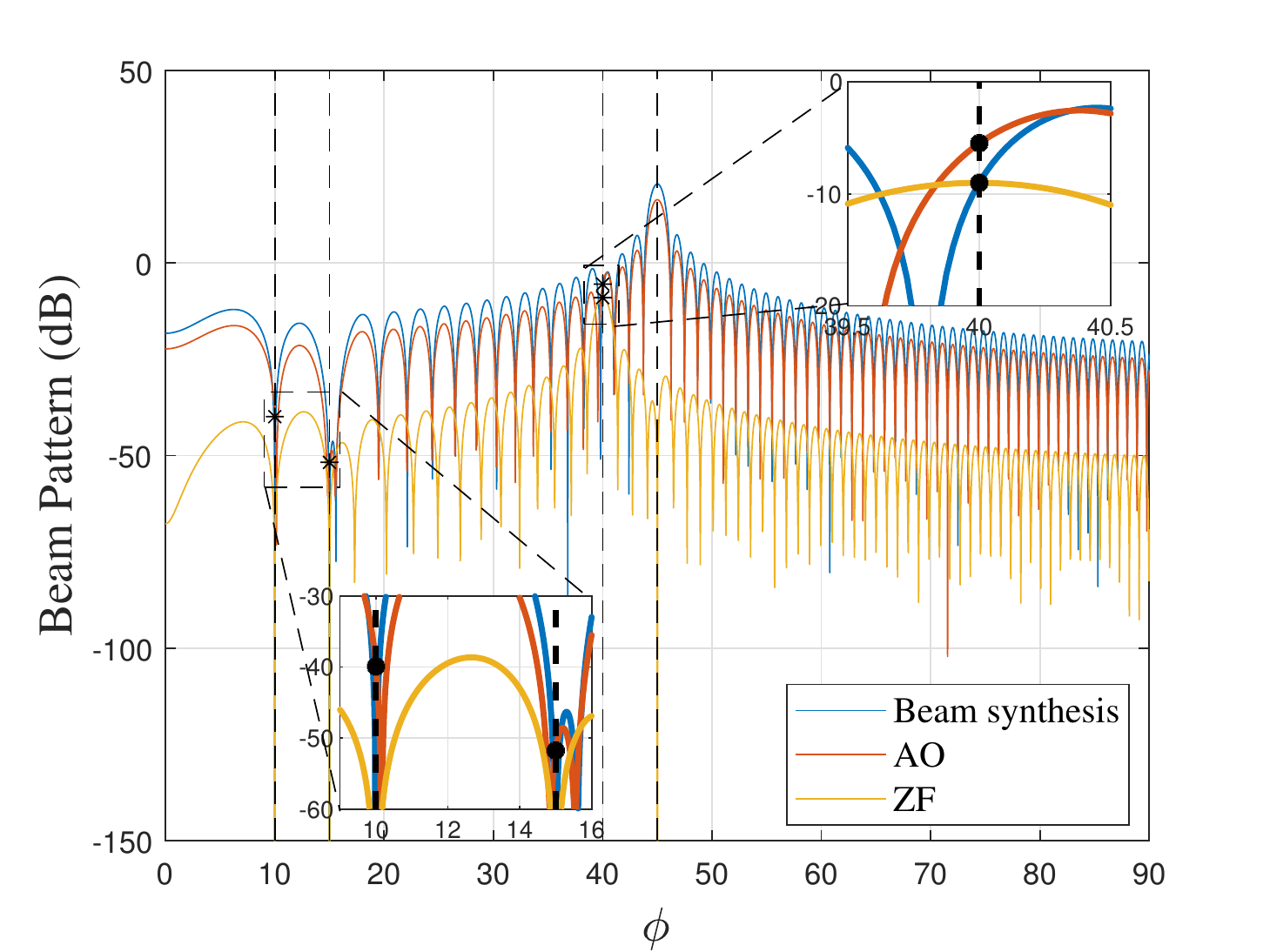}}\\
		\caption{Transmitted beam pattern for different data streams. (a) Data stream 1 for UE 1 ($10^\circ$); (b) Data stream 2 for UE 2 ($15^\circ$); (c) Data stream 3 for UE 3 ($40^\circ$).
		}
		\label{fig_beampattran0}
	\end{figure*}

	\textbf{Interference Management:} In Fig. \ref{fig_beampattran0}, we show how interference management and power allocation between sensing and communication are achieved by different transmitter structures, which further explains the performance difference shown in Fig. \ref{fig_BT}.	For ease of display, we fix the direction of the ST, the UEs and the clutter patches at $45^\circ$, $\{ 10^\circ,15^\circ,40^\circ  \}$ and $\{ 50^\circ,60^\circ \}$, respectively. Parts (a), (b) and (c) in Fig. \ref{fig_beampattran0} illustrate the beam pattern for each UE, together with their corresponding energy leakage to the ST. The beam pattern for the $i$th communication data stream is defined as $P_{i}(\phi)=||\mathbf{a}_T^\HH(\phi)\mathbf{f}_{c,i}||^2$. Here we set $\kappa_c=0.5$. 
	
	It can be observed that both ZF and B-syn eliminate the multi-user interference, but AO allows a low level of interference. In terms of the power leakage from communication to sensing, both B-syn and AO leak a certain amount of energy from the UEs to the ST, but ZF does not. Furthermore, with AO, the transmit gain on the ST direction from data streams 1, 2, and 3 are about $16.2145$dB, $16.2416$dB and $16.4111$dB, respectively, where the total gain is about $21.0612$dB. 
	The corresponding numbers for B-syn are $4.7362$dB, $11.0243$dB and $20.4923$dB, respectively, and the total gain is about $21.0598$dB. 
	Meanwhile, the resultant SCNR of AO and B-syn are $475.947$ and $472.001$, respectively.
	These observations agree with (\ref{qsolu0}), which indicates that $\beta_i$ is proportional to the $i$th entry of $\mathbf{a}_{tgt}$, i.e., $\mathbf{a}_T^\HH(\phi_t)\mathbf{f}_{ZF,i}$ and it is more efficient to leak energy to the ST from UEs closer to the ST (higher channel correlation).

	\begin{figure}[t]
		\centering
		\includegraphics[width=3.2in]{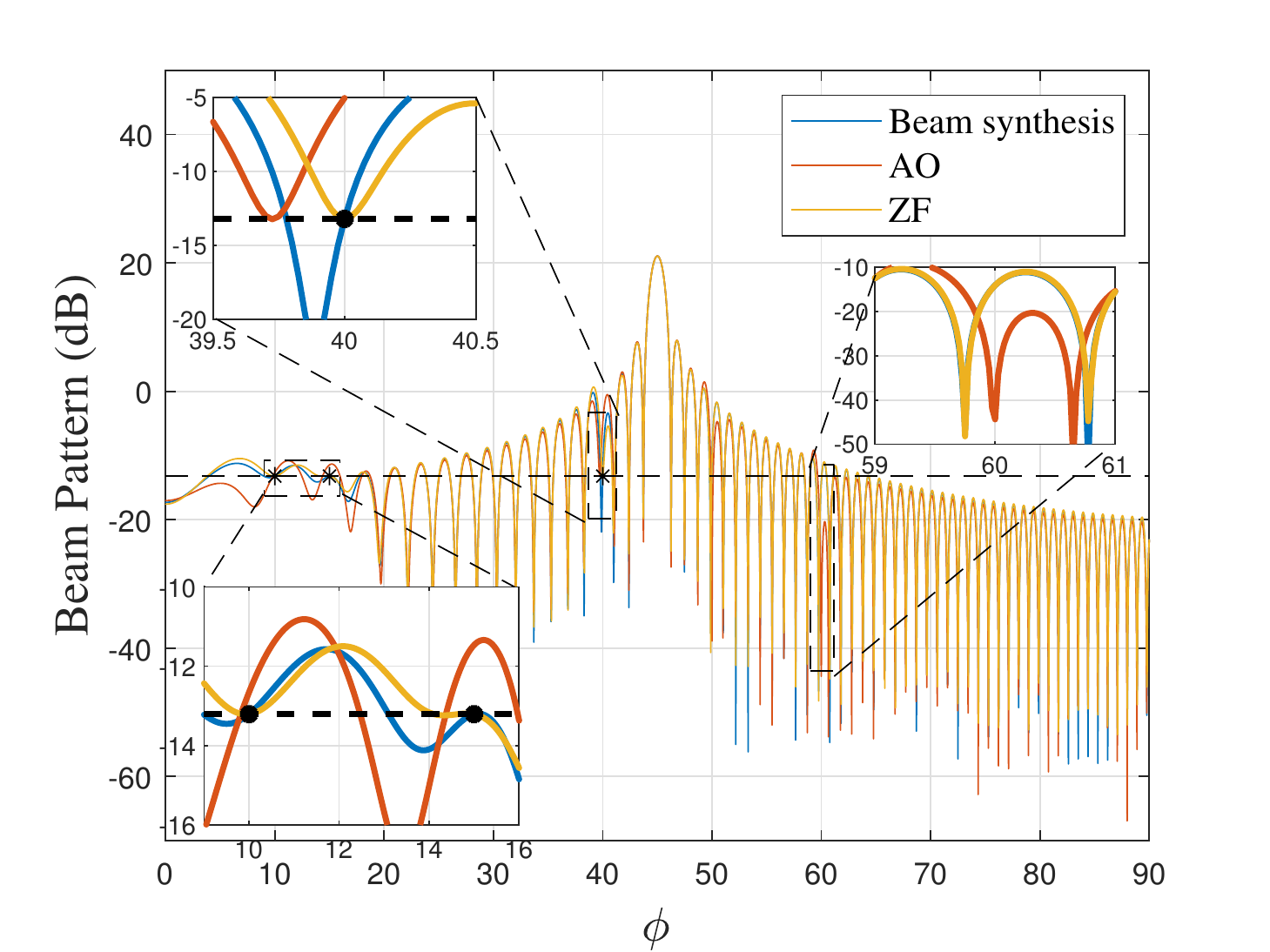}
		\caption{Overall Beam Pattern.}
		\label{fig_beampatternall}
	\end{figure}

	\textbf{Overall Beam Pattern:} Fig. \ref{fig_beampatternall} shows the overall beam pattern, i.e., $P(\phi) =||\mathbf{a}_T^\HH (\phi) \mathbf{F}||^2$.
	Comparing AO with B-syn, we can observe that they achieved the same communication performance by different strategies, i.e., AO delivers a higher power to the UEs while allowing interference between UEs, but B-syn forces the interference to zero while sending a lower power to different UEs. On the other hand, the transmit power towards the ST by two schemes is similar, which agrees with Fig. \ref{fig_BT}. When comparing B-syn with ZF, we notice that they achieved the same gain for the UEs but B-syn obtained a higher transmit power towards the ST, because leaking energy from communication to sensing is more efficient than forming a dedicated sensing signal, which agrees with (\ref{ineq}). Furthermore, we can observe that AO transmits extremely low (but non-zero) power on the direction of the clutter patches, i.e., $\{ 50^\circ,60^\circ \}$. As will be shown later, this will give more freedom to the receiver design.

	\begin{figure*}[t]
		\centering
		\subfloat[]{\includegraphics[width=3.2in]{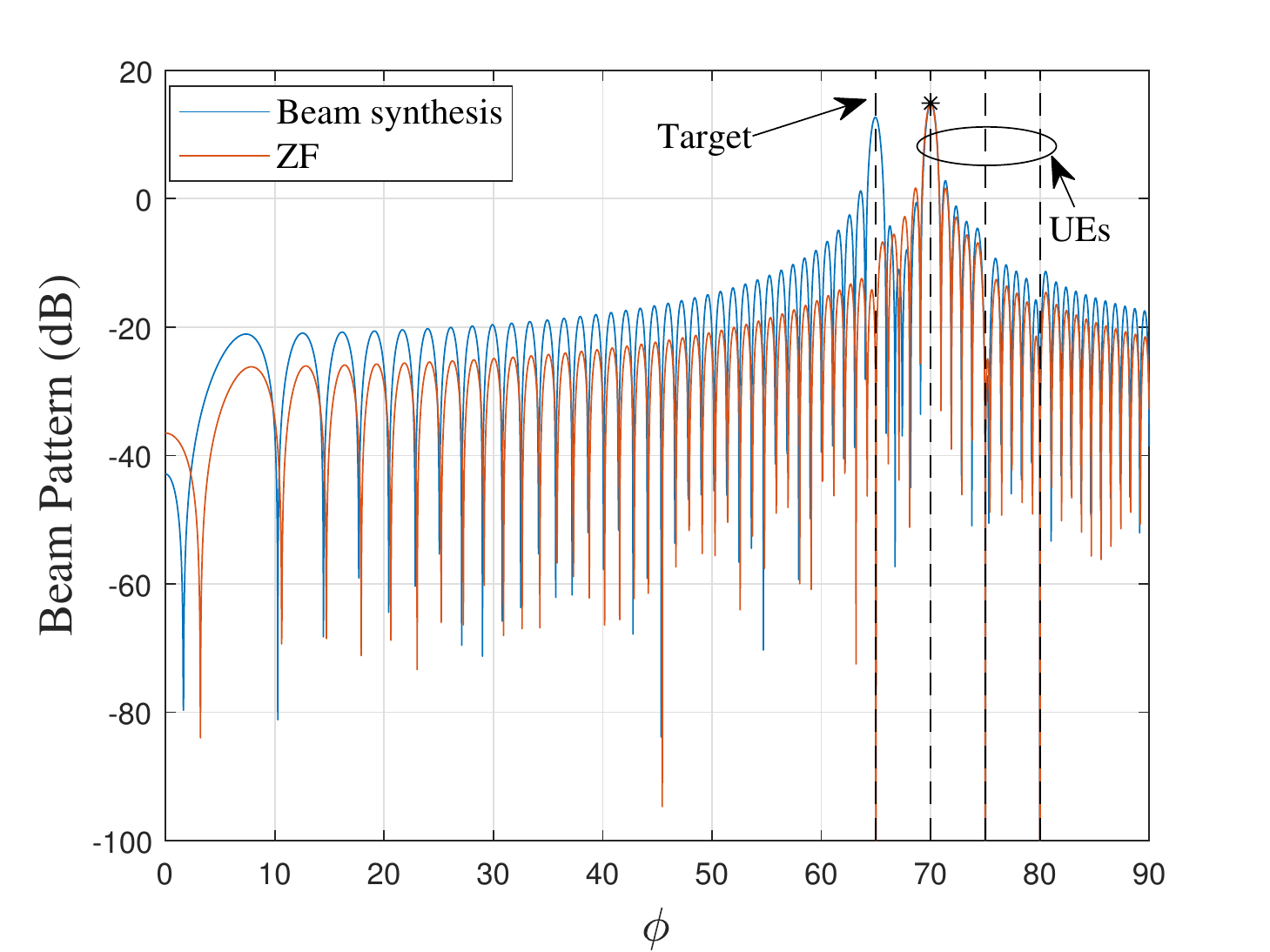}}\
		\subfloat[]{\includegraphics[width=3.2in]{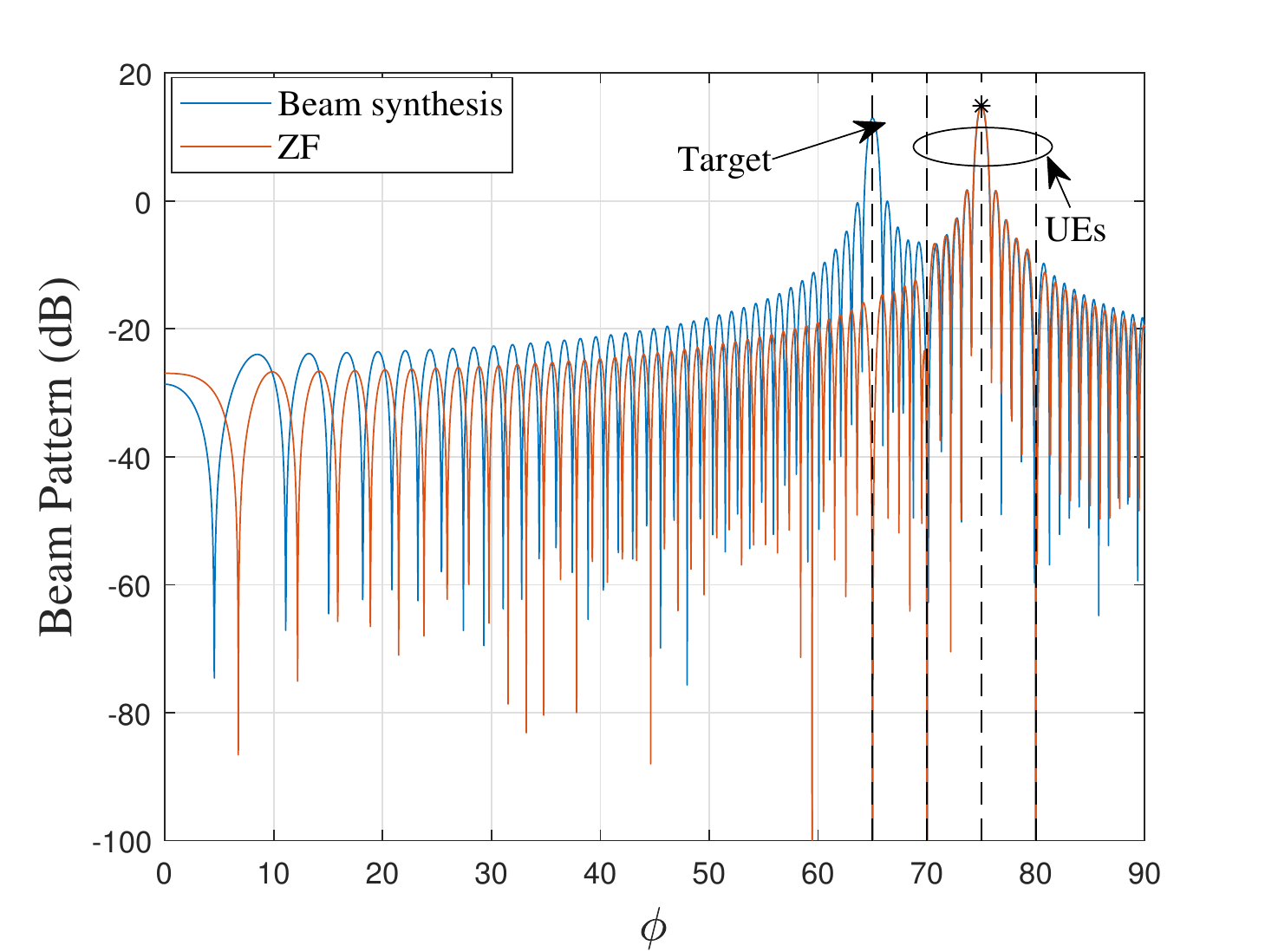}}\
		\subfloat[]{\includegraphics[width=3.2in]{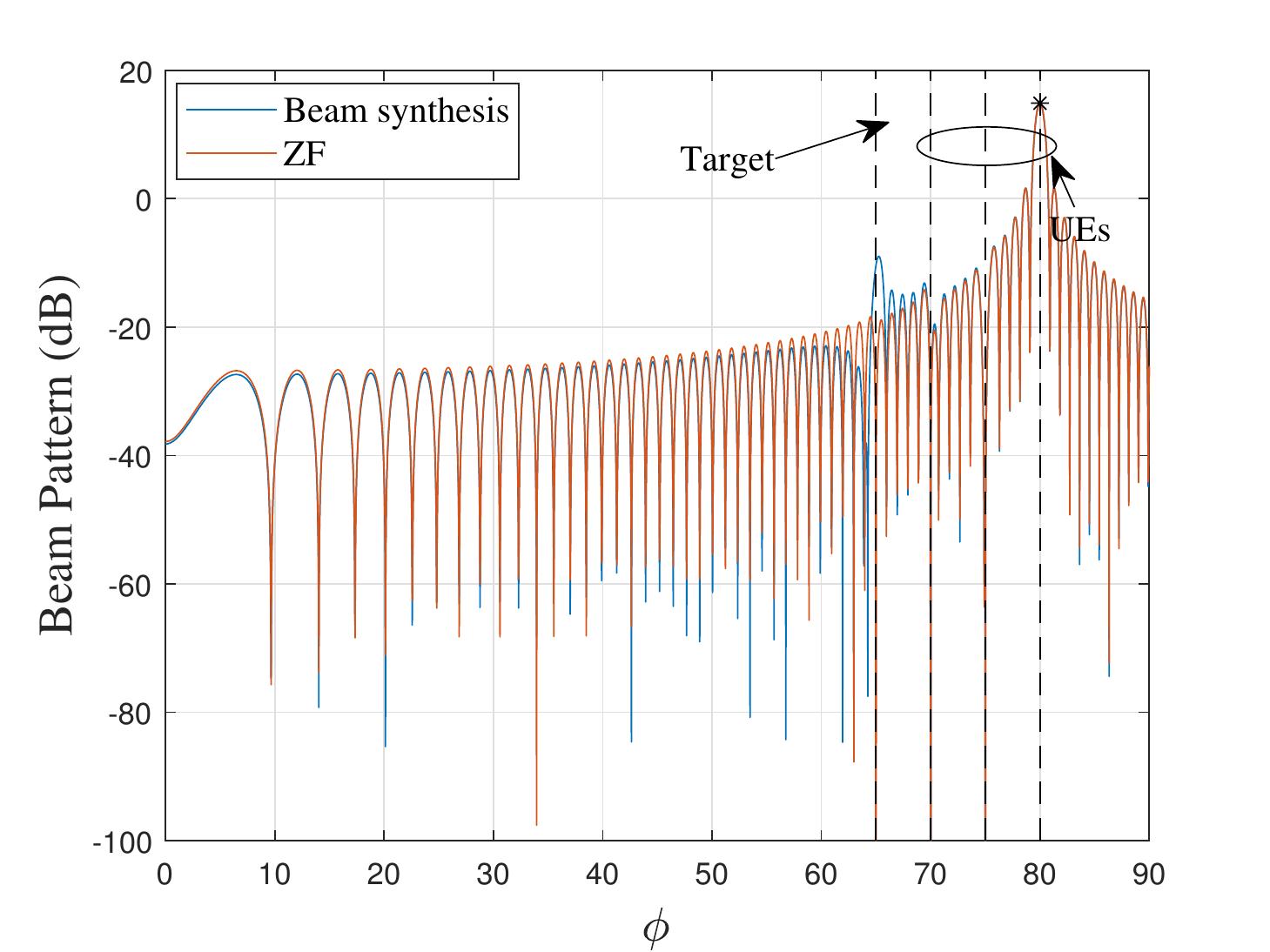}}\\
		\caption{Transmitted beam pattern for different location of ST and UEs. (a) $\phi_t=65^\circ$, Data stream 1 for UE 1 ($70^\circ$); (b) $\phi_t=65^\circ$,  Data stream 2 for UE 2 ($75^\circ$); (c) $\phi_t=65^\circ$, Data stream 3 for UE 3 ($80^\circ$).
		}
		\label{fig_beampattran1}
	\end{figure*}

	\textbf{Which UE Leaks More Energy?} Fig. \ref{fig_beampattran1} shows the beam pattern where the relative locations of the UEs with respect to the ST are different and we set  $\Gamma=600$. In general, the ST can obtain more gain from the closer UEs. The B-syn and ZF methods can always avoid multi-UE interference. 
	
	\begin{figure*}[t]
		\centering
		\subfloat[]{\includegraphics[width=3.2in]{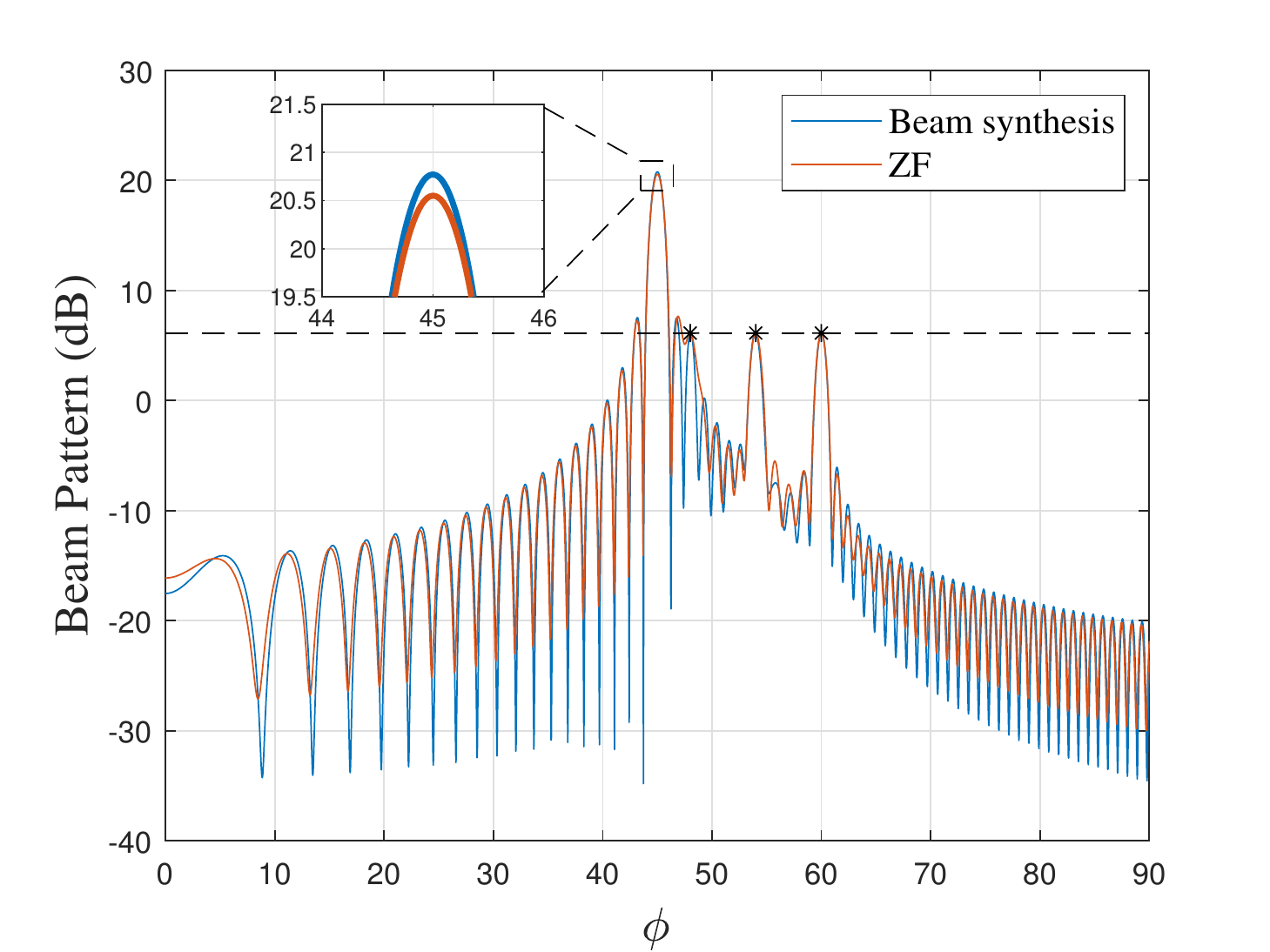}}\
		\subfloat[]{\includegraphics[width=3.2in]{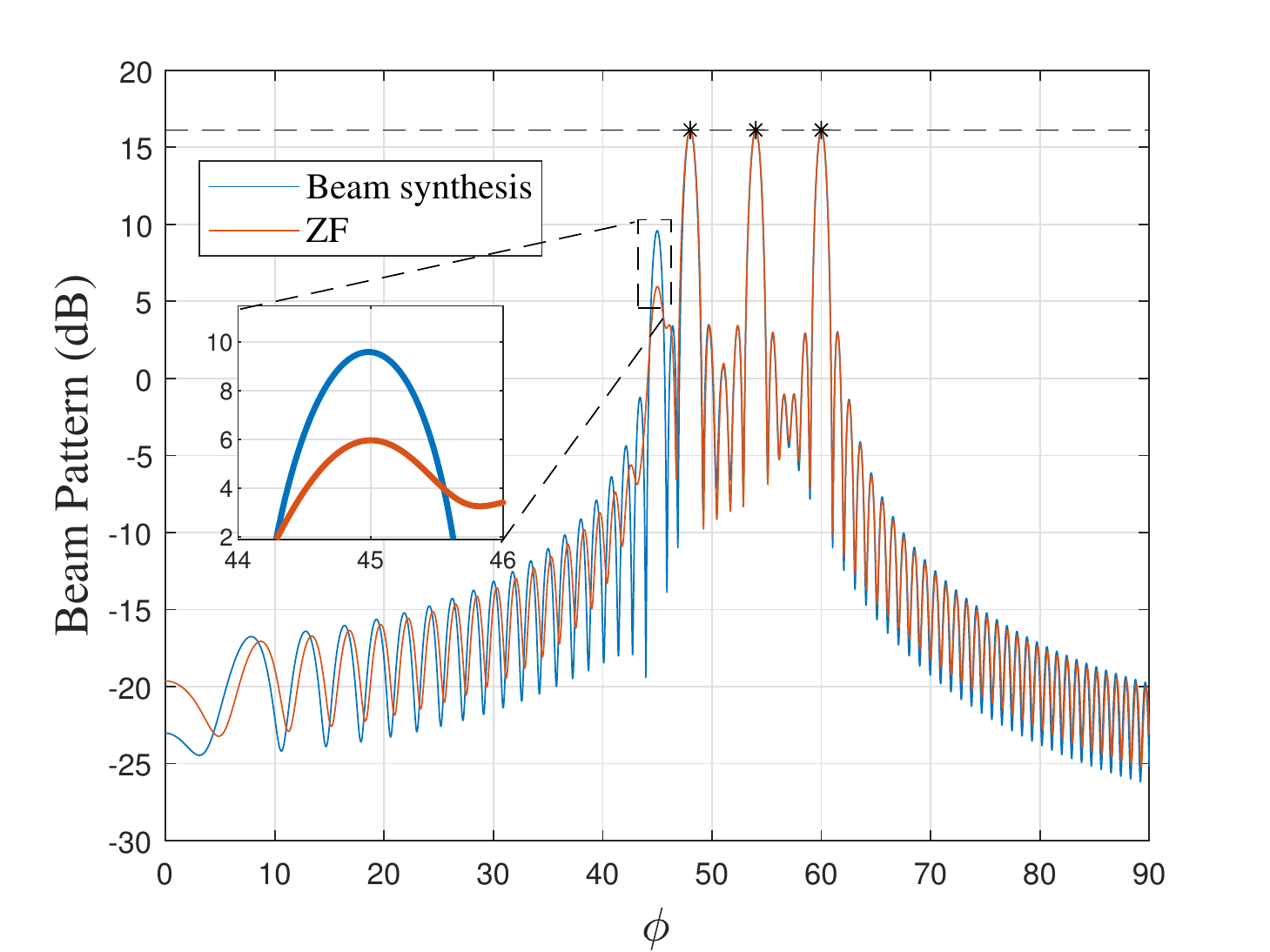}}\\
		\caption{Transmitted beam pattern under different $\Gamma$. (a) $\Gamma=80$; (b) $\Gamma=800$;.
		}
		\label{fig_beampattdiffGam}
	\end{figure*}
	
	\textbf{Impact of Communication Requirement:} Fig. \ref{fig_beampattdiffGam} shows the beam pattern with different $\Gamma$s. The directions of the ST and UEs are set as $45^\circ$ and $(48^\circ,54^\circ,60^\circ)$, respectively. The transmit power to the ST by B-syn is larger than that of ZF owing to  the power improvement in (\ref{ineq}).
	When $\Gamma$ is reasonably large, the gap  becomes larger. This is because, with a large $\Gamma$, the power improvement  will be more significant as more energy has been used for communication.

	\begin{figure*}[h]
		\centering
		\subfloat[]{\includegraphics[width=3.2in]{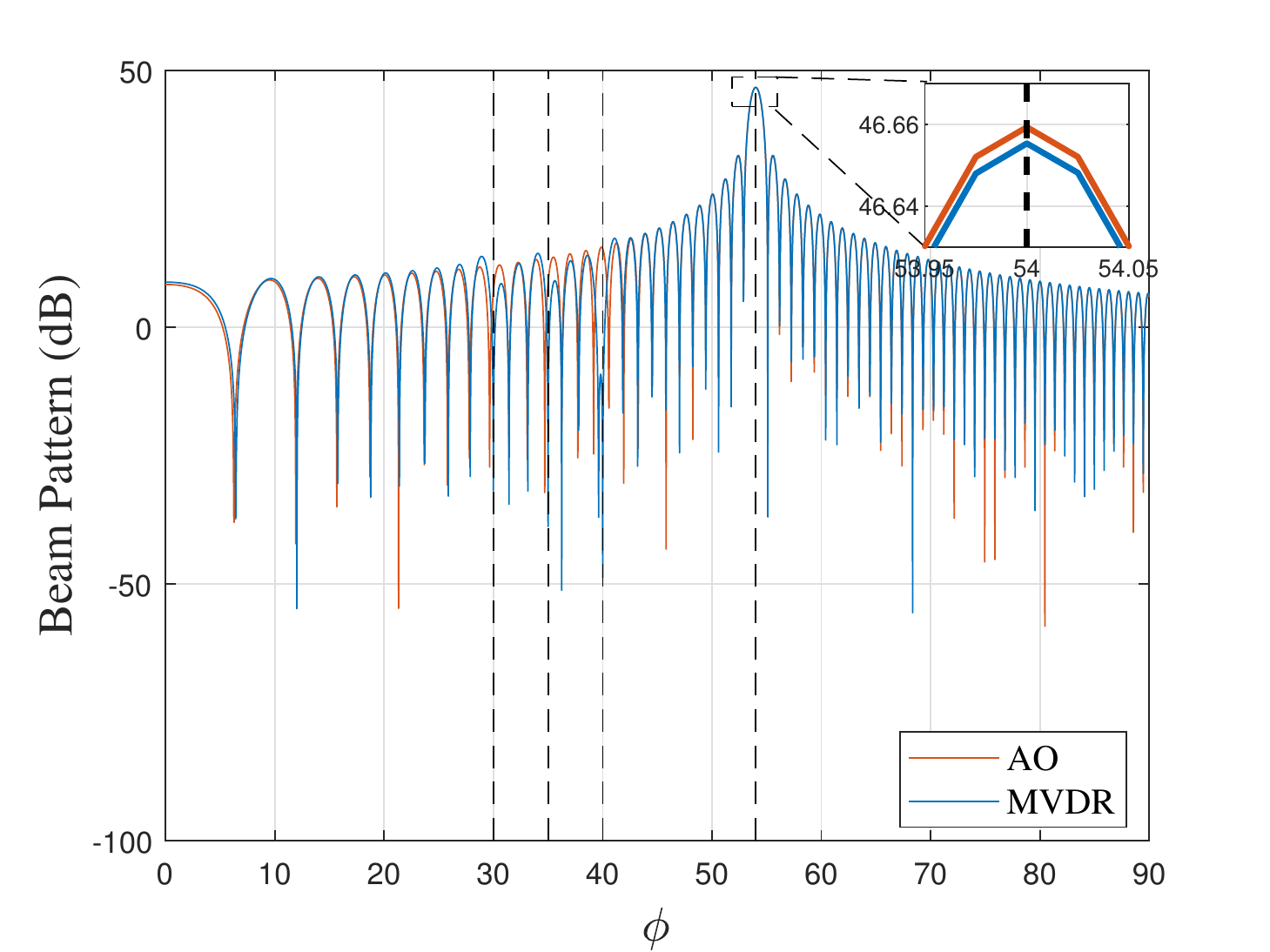}}\
		\subfloat[]{\includegraphics[width=3.2in]{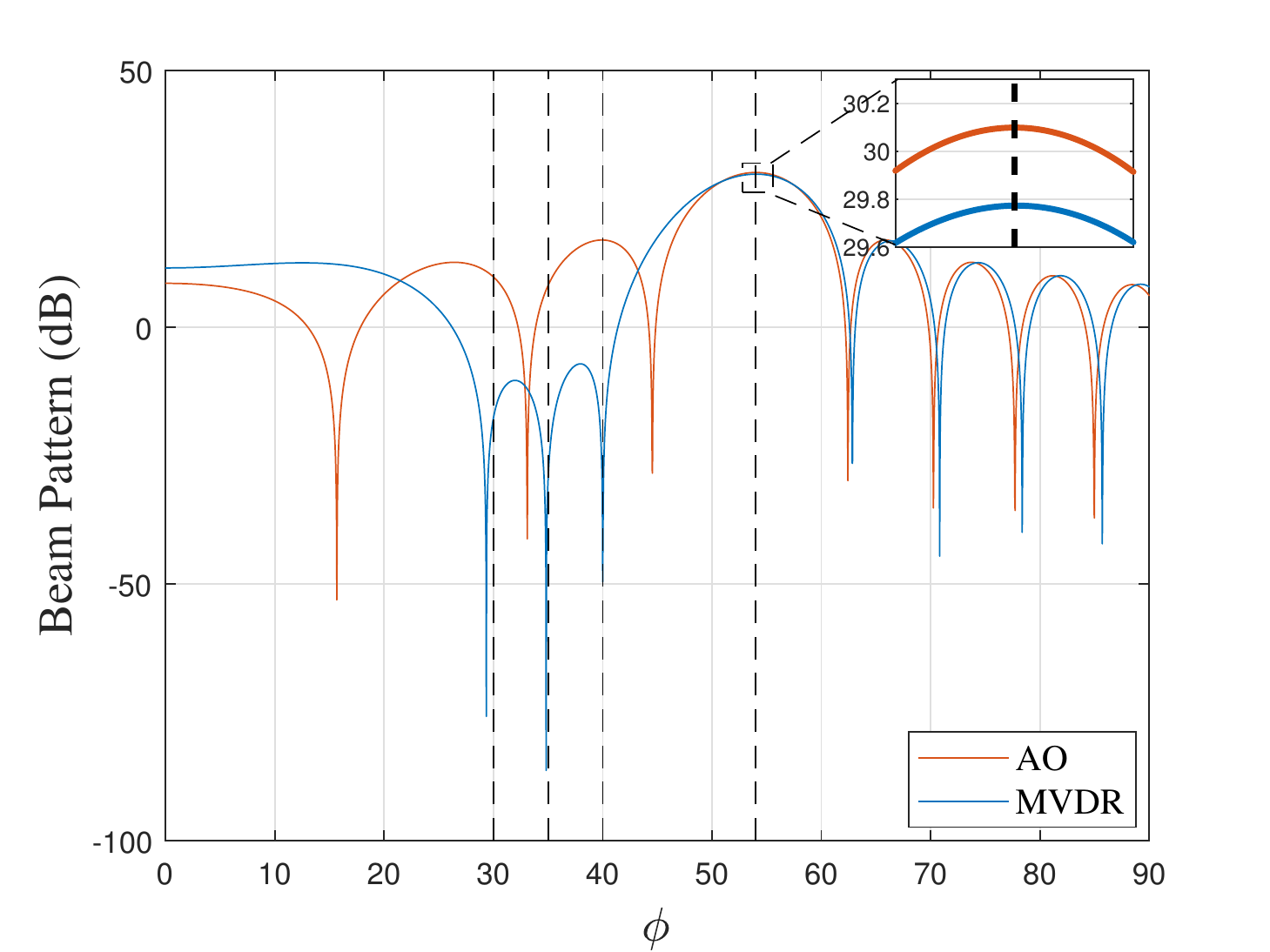}}\\
		\caption{Received beam pattern under different $N_r$. (a) $N_r=128$; (b) $N_r=16$;.
		}
		\label{fig_beampattreceive}
	\end{figure*}

	\textbf{Receive Beam Pattern:} Next, we show the beamforming performance of the TMT, i.e., $P(\phi)=||\mathbf{w}^\HH\mathbf{W}_{RF}^\HH\mathbf{a}_R(\phi)||^2$. For ease of illustration, we fix the direction of the ST and the clutter patches at $54^\circ$ and $(30^\circ,35^\circ,40^\circ)$, respectively. Fig. \ref{fig_beampattreceive} shows the received beam pattern with different  number of the receive antennas $N_r$. Overall, the mainlobe of all scenarios can focus on the ST while the responses on the clutter direction are all less than $-20$dB. Comparing  `B-syn $+$ MVDR' and AO, we can observe that the MVDR receiver suppresses the energy from the direction of clutter patches while AO does not need to. This is because AO can suppress the transmit power towards the clutter patches, which leaves more freedom for the receiver.
	
	\section{Conclusion}
	In this paper, we proposed a novel perceptive mobile network structure with distributed target monitoring terminals, which saves the full-duplex operation required by conventional ISAC systems. We investigate the interference management between sensing and communication by jointly design the transmit and receive beamformers to maximize the weighted average between sensing and communication performance. The problem was first solved by an AO method and linear transceiver structures were also derived to reduce the computation complexity and reveal  interesting physical insights. It was shown that leaking communication energy towards to the sensing target is more efficient than forming a dedicated sensing signal.  Furthermore, the amount of energy leakage depends on the channel correlation between the communication user and sensing target, which is determined by their locations. Simulation results validated the effectiveness of the proposed methods and illustrated the physical insights regarding interference management and resource allocation between sensing and communication.

	\appendices
	
	\section{Proof of Proposition \ref{PROconverge}}
	\label{IterConv}
	To simply the notation, we first denote 
	\begin{equation}
		\begin{split}\label{aurbur}
			\mathbf{a}_{\mathbf{u}_r}^{(t+1)}=\mathbf{F}^{(t+1),\HH}\mathbf{H}^{(t+1),\HH}\mathbf{W}_{RF}^{(t+1)}  \mathbf{w}^{(t+1)} ,\quad
			B_{\mathbf{u}_r}^{(t+1)}=\mathcal{P}_{Q}\left(\mathbf{w}^{(t+1)},\mathbf{W}_{RF}^{(t+1)},\mathbf{F}^{(t+1)} \right).
		\end{split}
	\end{equation}
	
	Hence, (\ref{urt1}) can be rewritten as $\mathbf{u}_r^{(t+1)}=\frac{\mathbf{a}_{\mathbf{u}_r}^{(t+1)}}{B_{\mathbf{u}_r}^{(t+1)}}$.
	By substituting (\ref{aurbur}) into (\ref{lossfunc}), we have
	\begin{equation}
		\begin{split} \nonumber
			&\mathcal{F}_R\left(\mathbf{u}_r^{(t)}|\mathbf{w}^{(t+1)},\mathbf{W}_{RF}^{(t+1)},\mathbf{F}^{(t+1)} \right)=2\kappa_r\Re(\mathbf{a}_{\mathbf{u}_r}^{(t+1),\HH} \mathbf{u}_r^{(t)})-\kappa_r \Vert \mathbf{u}_r \Vert^2 B_{\mathbf{u}_r}^{(t+1)}.\\
		\end{split}
	\end{equation}
	
	Thus, we can obtain
	\begin{equation}
		\begin{split}\nonumber
			&\mathcal{F}_R\left(\mathbf{u}_r^{(t+1)}|\mathbf{w}^{(t+1)},\mathbf{W}_{RF}^{(t+1)},\mathbf{F}^{(t+1)} \right)-\mathcal{F}_R\left(\mathbf{u}_r^{(t)}|\mathbf{w}^{(t+1)},\mathbf{W}_{RF}^{(t+1)},\mathbf{F}^{(t+1)} \right)\\
			&=\frac{\kappa_r}{B_{\mathbf{u}_r}^{(t+1)}}\left\Vert\mathbf{a}_{\mathbf{u}_r}^{(t+1)}-B_{\mathbf{u}_r}^{(t+1)} \mathbf{u}_r^{(t)} \right\Vert^2 \geq 0.
		\end{split}
	\end{equation}
	It indicates that the objective function $\mathcal{F}_R$ increases after the update of $\mathbf{u}_r$.
	Then we have
	\begin{equation}
		\begin{split}
			\label{SCNRinc}
			&\text{SCNR}^{(t+1)}=\text{SCNR}\left(\mathbf{w}^{(t+1)},\mathbf{W}_{RF}^{(t+1)},\mathbf{F}^{(t+1)} \right)=\frac{\mathcal{F}_R\left(\mathbf{w}^{(t+1)},\mathbf{W}_{RF}^{(t+1)},\mathbf{F}^{(t+1)},\mathbf{u}_r^{(t+1)} \right)}{\kappa_r}\\
			&\geq \frac{\mathcal{F}_R\left(\mathbf{w}^{(t+1)},\mathbf{W}_{RF}^{(t+1)},\mathbf{F}^{(t+1)},\mathbf{u}_r^{(t)} \right)}{\kappa_r}\geq \cdots \geq \frac{\mathcal{F}_R\left(\mathbf{w}^{(t)},\mathbf{W}_{RF}^{(t)},\mathbf{F}^{(t)},\mathbf{u}_r^{(t)} \right)}{\kappa_r}= \text{SCNR}^{(t)}.\\
		\end{split}
	\end{equation}
	
	Similarly, we have 
	\begin{equation}
		\begin{split}\nonumber
			&\mathcal{F}_k (u_k^{(t+1)}|\mathbf{F}^{(t+1)})- \mathcal{F}_k (u_k^{(t)}|\mathbf{F}^{(t+1)})= \frac{\kappa_c}{B_{u_k}^{(t+1)}}\left\Vert a_{u_{k}}^{(t+1)}-B_{u_{k}}^{(t+1)} u_k^{(t)} \right\Vert^2 \geq 0.
		\end{split}
	\end{equation}
	where $a_{u_{k}}^{(t+1)}=\mathbf{f}_c^{(t+1),\HH}\mathbf{h}_{c,k}^{(t+1)}$, $		B_{u_{k}}^{(t+1)}=	\sum_{i\neq k}\left|\mathbf{h}_{c,k}^\HH\mathbf{f}_{c,i} \right|^2
	+	\left\Vert\mathbf{h}_{c,k}^\HH\mathbf{F}_R \right\Vert^2
	+\sigma_c^2$.
	
	Denote $k^\dagger=\arg\min_{k \in [1,K]} \gamma_k (\mathbf{F}^{(t+1)})$. We have
	\begin{equation}
		\begin{split}
			\label{gammakinc}
			\min_{k}\gamma_k (\mathbf{F}^{(t+1)})
			=\frac{\mathcal{F}_{k^\dagger} (u_{k^\dagger}^{(t+1)}|\mathbf{F}^{(t+1)})}{\kappa_c}
			\geq \frac{\mathcal{F}_{k^\dagger} (u_{k^\dagger}^{(t)}|\mathbf{F}^{(t+1)})}{\kappa_c} \overset{(a)}{\geq} 
			\frac{\zeta^{(t)}}{\kappa_c}=\min_{k }\gamma_k (\mathbf{F}^{(t)}),
		\end{split}
	\end{equation}
	where (a) comes from the constraint in (\ref{probFt1}).
	
	From (\ref{SCNRinc}) and (\ref{gammakinc}), the sequence $\mathcal{L}^{(t)}=\kappa_R\text{SCNR}^{(t)}+\kappa_c \min_{k \in [1,K]}\gamma_k(\mathbf{F}^{(t)})$ is monotonically increasing with more iterations.
	According to the monotone convergence theorem \cite{bibby_1974}, the increasing sequence $\mathcal{L}^{(t)}$ will converge to a stationary point $\mathcal{L}^{\star}$ as $t$ increases. \hfill $\blacksquare$

	\section{Proof of Lemma \ref{mualemma}}
	\label{muaproof}
	For ease of illustration,  we denote $\mathbf{H}_e(\lambda_{a})$ as $\mathbf{H}_e$ hereafter in this proof.
	Recalling (\ref{Heisac}) and applying the block matrix inversion lemma \cite{9052470}, we can show that
	\begin{equation}
		\begin{split}
			(\mathbf{H}_e^\HH \mathbf{H}_e)^{-1}=\left[\begin{matrix}
				\mathbf{H}_c^\HH \mathbf{H}_c & \lambda_{a} \mathbf{H}_c^\HH \mathbf{a}_T(\phi_t)\\
				\lambda_{a} \mathbf{a}_T^\HH(\phi_t) \mathbf{H}_c   & \lambda_{a}^2 \\
			\end{matrix}\right]^{-1}=\left[\begin{matrix}
				(\mathbf{H}_c^\HH \mathbf{H}_c)^{-1} & \mathbf{0}\\
				\mathbf{0}^\HH  & 0 \\
			\end{matrix}\right]+\left[\begin{matrix}
				\mathbf{A}_1 & \mathbf{A}_2\\
				\mathbf{A}_2^\HH  & A_3^\HH \\
			\end{matrix}\right],
		\end{split}
	\end{equation}
	where
	\begin{equation}
		\begin{split}
			&\mathbf{A}_1=\frac{\lambda_{a}^2(\mathbf{H}_e^\HH \mathbf{H}_e)^{-1}\mathbf{H}_e^\HH\mathbf{a}_T(\phi_t)\mathbf{a}_T^\HH(\phi_t) \mathbf{H}_e(\mathbf{H}_e^\HH \mathbf{H}_e)^{-1}}{\lambda_{a}^2-\lambda_{a}^2\mathbf{a}_T^\HH(\phi_t) \mathbf{H}_e(\mathbf{H}_e^\HH \mathbf{H}_e)^{-1}\mathbf{H}_e^\HH\mathbf{a}_T(\phi_t)},\\
			&\mathbf{A}_2=\frac{\lambda_{a}(\mathbf{H}_e^\HH \mathbf{H}_e)^{-1}\mathbf{H}_e^\HH\mathbf{a}_T(\phi_t)}{\lambda_{a}^2-\lambda_{a}^2\mathbf{a}_T^\HH(\phi_t) \mathbf{H}_e(\mathbf{H}_e^\HH \mathbf{H}_e)^{-1}\mathbf{H}_e^\HH\mathbf{a}_T^(\phi_t)},\\
			&A_3=\frac{1}{\lambda_{a}^2-\lambda_{a}^2\mathbf{a}_T^\HH(\phi_t) \mathbf{H}_e(\mathbf{H}_e^\HH \mathbf{H}_e)^{-1}\mathbf{H}_e^\HH\mathbf{a}_T(\phi_t)}.\\
		\end{split}
	\end{equation}
	Thus, we have
	\begin{equation}
		\label{traceHehe}
		\begin{split}
			\tr(\mathbf{H}_e^\HH \mathbf{H}_e)^{-1}=\tr(\mathbf{H}_c^\HH \mathbf{H}_c)^{-1}+\tr(\mathbf{A}_1) + A_3.
		\end{split}
	\end{equation}
	By substituting (\ref{traceHehe}) into (\ref{mudef}), we have (\ref{mualammadef}). \hfill $\blacksquare$
	
		\section{Proof of Proposition \ref{Ptgt}}
	\label{ptgtproof}
%
	The transmit power to the ST is given by 
	\begin{equation}
		\begin{split}\nonumber
			&P_{\text{B-syn},tgt}\triangleq \mathbb{E}\left( ||\mathbf{a}_T^\HH(\phi_t) \mathbf{F}_{\text{B-syn}} \mathbf{s}||^2  \right)=\mathbb{E}\left(\left\Vert \sum_{i=1}^{K}\alpha_\star \mathbf{a}_T^\HH(\phi_t)\mathbf{f}_{\text{ZF},i}s_{c,i}+\sum_{i=1}^{K}\beta_{i} s_{c,i} +\sum_{j=1}^{N_s-K}\nu_{j} s_{R,j}\right\Vert^2 \right)\\
			&=  \sum_{i=1}^{K}\sum_{m=1}^{K}\alpha_\star^2 \mathbf{a}_T^\HH(\phi_t)\mathbf{f}_{\text{ZF},i} (\mathbf{a}_T^\HH(\phi_t)\mathbf{f}_{\text{ZF},m})^*  \mathbb{E}\left(s_{c,i}s_{c,m}^*\right) + \sum_{i=1}^{K}\sum_{m=1}^{K}\beta_{i}\beta_{m}^* \mathbb{E}\left(s_{c,i}s_{c,m}^*\right)\\
			& +\sum_{j=1}^{N_s-K}\sum_{n=1}^{N_s-K}\nu_{j}\nu_{n}^* \mathbb{E}\left(s_{R,j}s_{R,n}^*\right) +\sum_{i=1}^{K} \sum_{m=1}^{K} 2 \alpha_\star \Re\left( \beta_m^* \mathbf{a}_T^\HH(\phi_t)\mathbf{f}_{\text{ZF},i} \mathbb{E}\left(s_{c,i}s_{c,m}^*\right) \right) \\
			&+\sum_{i=1}^{K} \sum_{j=1}^{N_s-K} 2 \alpha_\star \Re\left( \nu_j^* \mathbf{a}_T^\HH(\phi_t)\mathbf{f}_{\text{ZF},i} \mathbb{E}\left(s_{c,i}s_{R,j}^*\right)\right) + \sum_{i=1}^{K} \sum_{j=1}^{N_s-K} 2\Re\left(\beta_{i}\nu_j^* \mathbb{E}\left(s_{c,i}s_{R,j}^*\right) \right),\\
		\end{split}
	\end{equation}
	where we utilized the property $\mathbf{a}_T^\HH \mathbf{f}_\perp=1$. Here,  $s_{c,i}$ and $s_{R,j}$ denote the $i$th and $j$th entry of $\mathbf{s}_c$ and $\mathbf{s}_R$, respectively.
	Recalling that $\mathbb{E}(\mathbf{s}\mathbf{s}^\HH)=\mathbf{I}$, we have
	\begin{equation}
		\label{ptgt}
		\begin{split}
			&P_{\text{B-syn},tgt}=\sum_{i=1}^{K} |\alpha_\star \mathbf{a}_T^\HH(\phi_t)\mathbf{f}_{\text{ZF},i} |^2+\sum_{i=1}^{K} 2\alpha_\star \Re( \beta_i^*\mathbf{a}_T^\HH(\phi_t)\mathbf{f}_{\text{ZF},i})  +\sum_{i=1}^{K} |\beta_i |^2+\sum_{j=1}^{N_s-K} |\nu_j|^2.\\
		\end{split}
	\end{equation}
	Next, we further simplify (\ref{ptgt}). First, by the definition of $\mathbf{F}_{\text{ZF}}$ and $\alpha_\star$, we have
	\begin{equation}
		\begin{split}
			\sum_{i=1}^{K} |\alpha_\star \mathbf{a}_T^\HH(\phi_t)\mathbf{f}_{\text{ZF},i} |^2
			&=\alpha_\star^2\mathbf{a}_T^\HH(\phi_t) \mathbf{F}_{\text{ZF}}\mathbf{F}_{\text{ZF}}^\HH \mathbf{a}_T(\phi_t)=\Gamma \sigma_c^2 \mathbf{a}_T^\HH(\phi_t) \mathbf{H}_{c}  \left(\mathbf{H}_{c}^\HH\mathbf{H}_{c}\right)^{-2}\mathbf{H}_{c}^\HH \mathbf{a}_T(\phi_t).
		\end{split}
		\label{aatfzf}
	\end{equation}
	
	Then, to meet the transmit power constraint, we have
	\begin{equation}\nonumber
		\begin{split}
			P&=\sum_{i=1}^{K}||\mathbf{f}_{\text{B-syn},c,i}||^2 + 	\sum_{j=1}^{N_s-K}||\mathbf{f}_{\text{B-syn},R,j}||^2=\sum_{i=1}^{K} \left(\alpha_\star^2 ||\mathbf{f}_{\text{ZF},i}||^2+|\beta_i|^2||\mathbf{f}_\perp||^2 \right)+\sum_{j=1}^{N_s-K} |\nu_j|^2 ||\mathbf{f}_\perp||^2,
		\end{split}
	\end{equation}
	which is equivalent to
	\begin{equation}
		\begin{split}
			\sum_{i=1}^{K} |\beta_i|^2+\sum_{j=1}^{N_s-K} |\nu_j|^2&=\frac{P- \alpha_\star^2 ||\mathbf{F}_{\text{ZF}}||^2}{||\mathbf{f}_\perp||^2}=(P-\Gamma \sigma_c^2 \tr(\mathbf{H}_c^\HH \mathbf{H}_c)^{-1})C_b,
		\end{split}
		\label{PowerCons}
	\end{equation}
	where we used the property $||\mathbf{F}_{\text{ZF}}||^2=P$ and $||\mathbf{f}_\perp||^2=\frac{1}{C_b}$. It requires that $P-\Gamma \sigma_c^2 \tr(\mathbf{H}_c^\HH \mathbf{H}_c)^{-1}\geq 0$, i.e., $\Gamma \leq \frac{P}{\sigma_c^2 \tr(\mathbf{H}_c^\HH \mathbf{H}_c)^{-1}}.$
	Substituting (\ref{PowerCons}) into (\ref{ptgt}) yields
		\begin{equation}\label{ptgt0}
		\begin{split}
			P_{\text{B-syn},tgt}
			&\triangleq \mathbb{E}\left(||a_T^\HH(\phi_t) \mathbf{F}_{\text{B-syn}}\mathbf{s}||^2 \right)
			=2\alpha_\star\Re(\mathbf{a}_{tgt}^\HH \mathbf{q}) + C_{tgt},
		\end{split}
	\end{equation}
	where 
	\begin{equation}
		\begin{split}
			\label{atgt}
			\mathbf{a}_{tgt}&=\left[\mathbf{f}_{\text{ZF},1}^\HH\mathbf{a}_T(\phi_t),\cdots,\mathbf{f}_{\text{ZF},K}^\HH\mathbf{a}_T(\phi_t),\underbrace{0,\cdots,0}_{N_s-K}\right]^\TT,\\
		\end{split}
	\end{equation}
	\begin{equation}\label{qdef}
		\begin{split}
			\mathbf{q}=[\beta_{1},\cdots,\beta_{K},\nu_1,\cdots,\nu_{N_s-K}]^\TT,
		\end{split}
	\end{equation}

		Omitting $\alpha_\star$ and $C_{tgt}$ which are constant irrelevant to the allocation of $\{\beta_{i}\}$ and $\{\nu_j\}$, the resource allocation problem to maximize $P_{\text{B-syn},tgt}$ is formulated  as
	\begin{equation}
		\begin{split}
			\max_{ \mathbf{q}}\quad &  \Re(\mathbf{a}_{tgt}^\HH \mathbf{q})\\
			s.t. \quad & ||\mathbf{q}||^2=P_q,
		\end{split}
		\label{preal0}
	\end{equation}
	where $P_q=(P-\Gamma \sigma_c^2 \tr(\mathbf{H}_c^\HH \mathbf{H}_c)^{-1})C_b$.
	The constraint is obtained from (\ref{PowerCons}) which indicates that the total power of $\mathbf{q}$, composed of $\{\beta_{i}\}_{i=1}^K$ and $\{\nu_j\}_j^{N_s-K}$, is fixed once $\Gamma$ is given. This problem	is to find a vector $\mathbf{q}$ on a sphere which has the highest correlation with $\mathbf{a}_{tgt}$. The solution can be obtained by the Lagrange multiplier method as
	\begin{equation}
		\label{qsolu0}
		\begin{split}
			\mathbf{q}=\sqrt{P_q}\cdot\frac{\mathbf{a}_{tgt}}{||\mathbf{a}_{tgt}||}.
		\end{split}
	\end{equation}
	
	It follows from (\ref{atgt}) and (\ref{qdef}) that $\nu_j=0$ for $j=1,\cdots,N_s-K$, which indicates that the optimal solution is to allocate all power to $\{\beta_i\}_{i=1}^{K}$. Substituting (\ref{qsolu0}) into (\ref{ptgt0}) yields   (\ref{ptgt1}). \hfill $\blacksquare$


\end{document}